\documentclass[11pt,a4paper]{article}
\pdfoutput=1 
\usepackage{graphicx}
\usepackage{caption}
\usepackage{subcaption}
\usepackage[utf8]{inputenc}

\usepackage[T1]{fontenc}

\usepackage{tikz,color}
\usetikzlibrary { decorations.pathmorphing, decorations.pathreplacing, decorations.shapes, }
\usepackage{array}

\usepackage{amsmath,amsthm,amssymb} 
\usepackage{dsfont,yfonts}
\usepackage{float}
\usepackage{array,xcolor,graphicx}
\usepackage{slashed}
\usepackage[normalem]{ulem}
\usepackage{jheppub}

\hypersetup{colorlinks=true,linkcolor=blue,citecolor=red,urlcolor=blue}

\newcommand{\Gammat}{\tilde{\Gamma}}

\definecolor{applegreen}{rgb}{0.0,0.7,0.0}

\def\im{\text{Im}\,}

\newcommand{\mn}{{\mu\nu}}
\newcommand{\ab}{{\alpha\beta}}

\date{\today}

\title{Spectra of correlators in the relaxation time approximation of kinetic theory}
 
\author[a]{Matej Bajec,}

\author[a,b]{Sa\v{s}o Grozdanov}
\author[a]{and Alexander Soloviev}

\affiliation[a]{Faculty of Mathematics and Physics, University of Ljubljana, Jadranska ulica 19, SI-1000 Ljubljana, Slovenia}
\affiliation[b]{Higgs Centre for Theoretical Physics, University of Edinburgh, Edinburgh, EH8 9YL, Scotland}

\abstract{
The relaxation time approximation (RTA) of the kinetic Boltzmann equation is likely the simplest window into the microscopic properties of collective real-time transport. Within this framework, we analytically compute all retarded two-point Green's functions of the energy-momentum tensor and a conserved $U(1)$ current in thermal states with classical massless particles (a `CFT') at non-zero density, and in the absence and presence of broken translational symmetry. This is done in $2+1$ and $3+1$ dimensions. RTA allows a full explicit analysis of the analytic structure of different correlators (poles versus branch cuts) and the transport properties that they imply (the thermoelectric conductivities, and the hydrodynamic, quasihydrodynamic and gapped mode dispersion relations). Our inherently weakly coupled analysis thereby also enables a direct comparison with previously known strongly coupled results in holographic CFTs dual to the Einstein-Maxwell-axion theories.
}

\begin{document}

\maketitle

\section{Introduction and summary of results}\label{sec:intro}

Microscopic dynamics of the `fundamental' constituents of matter gives rise to collective, macroscopic dynamics, including to fluids and gases. Such collective behavior, however, depends on complicated dynamics of many particles: for example, there are on the order of $10^{25}$ molecules in a liter of water. Whether these constituents (particles, atoms, molecules) are classical or quantum, a complete description of such a system at the level of each particle is intractable. Instead, the simplest physically transparent description of such dynamics combines methods of statistics and detailed knowledge of individual collisions. Such a description of a many-body system is known as kinetic theory and it crucially relies on the concept of quasiparticles (see e.g.~\cite{grad-1963,gross-1959,liboff-book,chapman-book,boltzmann-book,ford-book,silin-book,ferziger-kaper-book}). A way to understand the full, time-reversal-invariant microscopic dynamic in kinetic theory is the so-called BBGKY hierarchy \cite{yvon,bogoliubov,kirkwood,borngreen}, which couples together the dynamics of all $n$-particle distribution functions over the phase space. The famous Boltzmann equation is a truncation of this infinite chain of integro-differential equation to a single dynamical equation for a one-particle distribution function $f(t, {\bf x}, {\bf p})$ in the phase space that depends on the {\em collision integral} or the {\em collision kernel} accounting for the microscopic dynamics of 2-to-2 particle collisions. Even then, analyzing the Boltzmann equation exactly and solving it even numerically remains extremely difficult. For recent progress in finding such solutions in certain cases, see Refs.~\cite{Bazow:2015dha,Moore:2018mma,Grozdanov:2018atb,Denicol:2022bsq,Ochsenfeld:2023wxz}.

The Boltzmann equation can be further reduced to a simple differential equation within the context of the so-called {\em relaxation time approximation} (RTA), also known as the Bhatnagar-Gross-Krook approximation \cite{PhysRev.94.511}. One way to understand this approximation, in physical terms, is by considering the possible spectrum of the linearized collision integral (see Ref.~\cite{Grozdanov:2016vgg}). If we assume that an eigenvalue exists that is `parametrically' separated from the remaining ones, which dominates the late-time relaxation of the system with a single relaxation time $\tau_R$, then we can approximate the collision integral by the well-known RTA approximation \cite{anderson} discussed in detail below. Within this approximation, the Boltzmann equation can often be solved analytically. While in itself an `effective' and truncated (UV-incomplete) description, such explicit results are extremely instructive and valuable for the analysis of (weakly-coupled) low-energy collective states of matter (see e.g.~Refs.~\cite{Jaiswal:2013npa,Romatschke:2015gic,Denicol:2011fa,Romatschke:2017ejr,Rocha:2021zcw,Amoretti:2023hpb,Martinoia:2024cbw} and references therein). 

While often used outside of its appropriate regime in the literature, the kinetic description fundamentally relies on the weakly-coupled picture of a dilute gas of particles colliding with each other. This intrinsic limitation of such a description therefore means that if the particles become strongly interacting, it is hard to justify the numerous limits of the underlying quantum field theory (QFT) that we routinely make to render kinetic calculations tractable. An independent approach to the study of strongly coupled states of matter that is microscopic and UV-complete is the holographic duality (the AdS/CFT correspondence) \cite{Maldacena:1997re} (see Refs.~\cite{jorge-book,zaanen-book,Ammon:2015wua,hartnoll2018holographic}). Its limitation, on the other hand, is that it is in practice limited to large-$N$, typically supersymmetric conformal theories (CFTs). Nevertheless, holography has provided us with a wealth of invaluable information about thermal and dense collective states of matter. Despite the formidable difficulties in using holography beyond the large-$N$ limit (see Refs.~\cite{Denef:2009yy} and \cite{hartnoll2018holographic}), computations beyond infinite 't Hooft coupling can be performed by the inclusion of classical higher-derivative supergravity corrections to the bulk Einstein-Hilbert action in $4+1$ dimensions \cite{Gubser:1998nz} (see also \cite{Buchel:2004di,Grozdanov:2016vgg}). As per the holographic description, spectra of correlation functions correspond to the ring-down spectra of dual linearly perturbed black holes: the quasinormal modes \cite{Kovtun:2005ev}. In a sense, quasiparticle excitations in kinetic theory spectra are replaced by the quasinormal modes of black holes. The central feature of the holographic finite-temperature ($T\neq 0 $) spectra are isolated simple poles and the absence of branch cuts, which were found in a finite-$T$ QFT calculation at zero coupling by Hartnoll and Kumar \cite{Hartnoll:2005ju}. The transition from one picture to another, as a function of the coupling constant, has been relatively well understood in recent year from the point of view of holography \cite{Waeber:2015oka,Grozdanov:2016vgg,Grozdanov:2016fkt,Solana:2018pbk,Grozdanov:2018gfx} (for related earlier work on current correlators in finite density states in $2+1$ dimensions, see Refs.~\cite{Witczak-Krempa:2012qgh,Witczak-Krempa:2013xlz}). Studies of spectra from the angle of finite-$T$ are unfortunately dramatically harder and are therefore extremely rare, existing only in the simple large-$N$ vector models \cite{Romatschke:2022jqg,Romatschke:2023sce}.

A natural question that arises in the investigation of thermal states and their physical excitations is what can be learnt from the simplification of weakly-coupled QFTs: namely, from kinetic theory. Despite the fact that the Boltzmann equation was written down in its full form in 1872 and has been extensively studied since then, it was only in 2015 that Romatschke analyzed the full analytic structure of retarded two-point correlation functions in RTA \cite{Romatschke:2015gic}. For subsequent extensions, see Refs.~\cite{Kurkela:2017xis,Kurkela:2019kip}. This $3+1$ dimensional calculation could then be compared with the spectra obtained from the holographic bulk Einstein gravity with the thermal Schwarzschild black brane background solution in $4+1$ dimensions and its finite coupling extensions (cf.~Ref.~\cite{Grozdanov:2016vgg}). What the RTA kinetic theory analysis of \cite{Romatschke:2015gic} shows is the combination of a branch cut along with the `universal' hydrodynamic poles present in any finite-$T$ holographic calculation. For discussions of linearized hydrodynamics, hydrodynamic correlators and the nature of dispersion relations of diffusion and sound, see Refs.~\cite{Kovtun:2012rj,Glorioso:2018wxw,Grozdanov:2019uhi}. For a discussion of the convergence of the hydrodynamic modes in RTA, see ref.~\cite{Heller:2020hnq}. The full picture of how QFTs interpolate between the weakly-coupled picture with branch cuts and the strongly-coupled picture with only poles remains to be understood although a highly suggestive and potentially complete mechanism has emerged from the holographic studies of higher-derivative gravity \cite{Grozdanov:2016vgg,Grozdanov:2018gfx}.

Holography has enabled us to study not only thermal states but, among numerous other setups, also finite-density states with a finite chemical potential $\mu \neq 0$. The easiest way to incorporate a $U(1)$ charge into the boundary CFT is to study the bulk Einstein-Maxwell theory with the Reissner-N\"{o}rdstrom black brane background. For holographic analyses of this theory, see Refs.~\cite{Son:2006em,Myers:2007we,Edalati:2010hk,Davison:2011uk,Amoretti:2014zha,Jansen:2020hfd,Abbasi:2020ykq}. On the other hand, holographic models seeking closer contact with `realistic' condensed matter systems that include a lattice and, for example, give rise to a finite conductivity in a $\mu \neq 0$ state, were developed by explicitly breaking the boundary translational symmetry in Refs.~\cite{Andrade:2013gsa,Vegh:2013sk,Davison:2013jba}. In particular, the simplest of these models, the so-called Einstein-Maxwell-dilaton model which accounts for finite $T$, $\mu$ and a translation-symmetry breaking parameter $\Gamma$ was then studied in detail by Davison and Gout\'{e}raux \cite{Davison:2014lua}. Some other papers studying holography with broken translations include Refs.~\cite{Davison:2015bea,Blake:2018leo,Baggioli:2021xuv,Baggioli:2021ejg,Baggioli:2022uqb,Ahn:2024aiw}. Finite coupling corrections to the picture they developed are presently not well understood.  

The goal of this paper is to bridge the gap between a weakly coupled description of matter in terms of kinetic theory and the strongly coupled holographic studies of thermal states that exhibit some combination finite density and broken translations. We analytically compute {\em all} retarded two-point functions of conserved operators in such systems by developing and solving in each case the appropriate RTA kinetic theory equations. This then allows us to classify and study the behaviour of physical modes that dominate the low-energy (late-time) dynamics of collective states, which are controlled by the conservation laws of operators corresponding to continuous global symmetries: the energy-momentum tensor $T^{\mu\nu}$ and the $U(1)$ current $J^{\mu}$. These conservation laws (or the Ward identities) are 
\begin{align}\label{eq:MomentumBreak}
\nabla_\mu T^{\mu\nu} &= \begin{cases}
0, & \nu\neq i,\\
- \Gamma T^{0i}, & \nu=i,
\end{cases} \\
\nabla_\mu J^\mu &= 0\label{eq:ChargeConservation},
\end{align}
where we have allowed for the breaking of momentum conservation with the help of the parameter $\Gamma$. In each case, we derive the full set of dispersion relation of all hydrodynamic and gapped modes visible by RTA as well as various transport coefficients, such as the {\em thermoelectric conductivities} accessible by the linear response theory and the Kubo formulae. A number of examples also exhibits interesting interplay between hydrodynamic and low-energy gapped modes (i.e., their `collisions' in the complex frequency plane), which can be understood in the context of {\em quasihydrodynamics} when the pole collisions occur near the hydrodynamic regime \cite{Grozdanov:2018fic}. The complete classification of all cases computed from RTA in $3+1$ and $2+1$ dimensions is collected in Tables \ref{table-3plus1} and \ref{table-2plus1}, respectively. 

More precisely, in these two tables, we {\em schematically} depict the qualitative properties of the RTA spectra of all channels of two-point functions composed of $T^{\mu\nu}$ and $J^\mu$ in the complex frequency $\omega$ plane for some fixed (small or on the order comparable to other scales in the problem) wavevector (or `momentum') $k$ aligned with the $x$ axis. In $3+1$ dimensions, the remaining symmetries allow for a decomposition of correlators into decoupled spin 0, 1 and 2 channels with respect to the $SO(2)$ symmetry of spatial rotations about the momentum vector. In $2+1$, they can be decomposed into even and odd channels with respect to the $\mathbb{Z}_2$ symmetry associated with parity transformations about the momentum vector. 

Each spectrum in $3+1$ and $2+1$ dimensions contains a branch cut running between the branch points at $\omega = \pm k - i /\tau_R$, where $\tau_R$ is the RTA relaxation time. Most spectra also contain hydrodynamic modes (gapless modes when $\Gamma = 0$), which we represent with black dots. Diffusive modes are placed on the imaginary axis and a pair of sound modes is placed symmetrically (with respect to the imaginary axis) in the complex plane at the same imaginary value as the diffusive poles. Some spectra also contain gapped modes. A gapped mode independent of the existence of momentum relaxation (represented with a green dot) exists only in $2+1$ dimensional even current-current (density-density) correlators and exhibits a collision with the diffusive mode after which both modes become propagating. This phenomenon is represented by blue arrows. It marks an important qualitative difference between the behavior of current correlators in $2+1$ and $3+1$ dimensions. We interpret this $2+1$ dimensional result to be analogous to the collision of poles associated with a `hydrodynamic-to-relativistic' crossover that was described in the context of holography at infinite coupling in Ref.~\cite{Witczak-Krempa:2013xlz}. In that case, as $k$ increases from $k=0$, the spectrum of quasinormal modes, which all lie on the imaginary $\omega$ axis, exhibits sequential pole collisions after which the poles move off the axis (obtaining a real part) in a `zipper-like' manner (see an earlier review \cite{Berti:2009kk} and a recent discussion of this phenomenon from the point of view of S-duality specific to $2+1$ dimensional CFTs in Ref.~\cite{Grozdanov:2024wgo}). The second type of gapped modes exists as a result of momentum relaxation (plotted with a black dots) and exhibits quasihydrodynamic collisions with a diffusive mode. Those collisions are represented by red arrows. The $2+1$ dimensional $\langle JJ \rangle$ correlator with momentum relaxation parallel to momentum is special in that it exhibits both types of pole collisions. Finally, we note that at $\mu = 0$, the $\langle T^{\mu\nu} J^\rho \rangle$ correlators vanish.    

\begin{table}[h!]
    \centering
    \resizebox{\columnwidth}{!}{%

    \begin{tabular}{c|c|c|c|c|c|c|c|c|}
         &
       \multicolumn{2}{c|}{
         $\langle J J \rangle$}
         &
         \multicolumn{3}{c|}{
         $\langle TT \rangle$}
         &
         \multicolumn{2}{c|}{
         $\langle TJ \rangle$}
       \\
       \hline
          & spin 0 & spin 1
         & spin 0 & spin 1 & spin 2 & spin 0
         & spin 1
         \\
         \hline
          $T\neq 0$ & 
\begin{tikzpicture}
\draw[ gray, thick] (-0.75,0)--(0.75,0) 
;
\draw[decorate,decoration={zigzag,segment length=4pt}](-0.5,-0.5) -- (0.5,-0.5);
\draw[ gray, thick] (0,-1.08)--(0,0.2) 
;
\foreach \Point in {(0.0,-0.2)}{
    \node at \Point {\textbullet};
}
\end{tikzpicture}
&
\begin{tikzpicture}
\draw[help lines, color=gray!30, dashed] 
;
\draw[gray, thick] (-0.75,0)--(0.75,0) 
;
\draw[decorate,decoration={zigzag,segment length=4pt}](-0.5,-0.5) -- (0.5,-0.5);
\draw[ gray, thick] (0,-1.08)--(0,0.2) 
;
\end{tikzpicture}
         & 
\begin{tikzpicture}
\draw[help lines, color=gray!30, dashed] 
;
\draw[ gray, thick] (-0.75,0)--(0.75,0) 
;
\draw[decorate,decoration={zigzag,segment length=4pt}](-0.5,-0.5) -- (0.5,-0.5);
\draw[ gray, thick] (0,-1.08)--(0,0.2) 
;
\foreach \Point in {(-0.3,-0.2), (0.3,-0.2)}{
    \node at \Point {\textbullet};
}
\end{tikzpicture}
&
\begin{tikzpicture}
\draw[help lines, color=gray!30, dashed] 
;
\draw[ gray, thick] (-0.75,0)--(0.75,0) 
;
\draw[decorate,decoration={zigzag,segment length=4pt}](-0.5,-0.5) -- (0.5,-0.5);
\draw[ gray, thick] (0,-1.08)--(0,0.2) 
;
\foreach \Point in {(0.0,-0.2)}{
    \node at \Point {\textbullet};
}
\end{tikzpicture}
&
\begin{tikzpicture}
\draw[ gray, thick] (-0.75,0)--(0.75,0) 
;
\draw[decorate,decoration={zigzag,segment length=4pt}](-0.5,-0.5) -- (0.5,-0.5);
\draw[ gray, thick] (0,-1.08)--(0,0.2) 
;
\end{tikzpicture}
&
---
& 
---
\\
        \hline
         $T, \Gamma_\parallel\neq0$  &\begin{tikzpicture}
\draw[ gray, thick] (-0.75,0)--(0.75,0) 
;
\draw[decorate,decoration={zigzag,segment length=4pt}](-0.5,-0.5) -- (0.5,-0.5);
\draw[ gray, thick] (0,-1.08)--(0,0.2) 
;
\foreach \Point in {(0.0,-0.2)}{
    \node at \Point {\textbullet};
}
\end{tikzpicture}
&
\begin{tikzpicture}
\draw[help lines, color=gray!30, dashed] 
;
\draw[ gray, thick] (-0.75,0)--(0.75,0) 
;
\draw[decorate,decoration={zigzag,segment length=4pt}](-0.5,-0.5) -- (0.5,-0.5);
\draw[ gray, thick] (0,-1.08)--(0,0.2) 
;
\end{tikzpicture}
&

\begin{tikzpicture}
\draw[gray, thick] (-0.75,0)--(0.75,0);
\draw[gray, thick] (0,-1.08)--(0,0.2);
\draw[->,red, thick] (0.0,-0.35)--(-0.3,-0.35);
\draw[->,red, thick] (-0.0,-0.35)--(0.3,-0.35);
\draw[->_,red, thick] (-0.0,-0.2)--(0.0,-0.35) ;
\draw[->_,red, thick] (-0.0,-0.6)--(0.0,-0.35) ;
\draw[decorate,decoration={zigzag,segment length=4pt}] (-0.5,-0.7) -- (0.5,-0.7);
\node at (-0.0,-0.5) [circle,black,fill,inner sep=1.5pt]{};
\node at (0.0,-0.2) [circle,black,fill,inner sep=1.5pt]{};
\end{tikzpicture}

&
\begin{tikzpicture}
\draw[gray, thick] (-0.75,0)--(0.75,0) 
;
\draw[decorate,decoration={zigzag,segment length=4pt}] (-0.5,-0.7) -- (0.5,-0.7);
\draw[gray, thick] (0,-1.08)--(0,0.2) 
;
\foreach \Point in {(-0.0,-0.2)}{
    \node at \Point {\textbullet};
}
\end{tikzpicture}
&

\begin{tikzpicture}
\draw[gray, thick] (-0.75,0)--(0.75,0) 
;
\draw[decorate,decoration={zigzag,segment length=4pt}] (-0.5,-0.7) -- (0.5,-0.7);
\draw[gray, thick] (0,-1.08)--(0,0.2) 
;

\end{tikzpicture}
&---
&---
         \\

         \hline
         $T, \Gamma_\perp\neq0$  &\begin{tikzpicture}
\draw[ gray, thick] (-0.75,0)--(0.75,0) 
;
\draw[decorate,decoration={zigzag,segment length=4pt}](-0.5,-0.5) -- (0.5,-0.5);
\draw[ gray, thick] (0,-1.08)--(0,0.2) 
;
\foreach \Point in {(0.0,-0.2)}{
    \node at \Point {\textbullet};
}
\end{tikzpicture}
&
\begin{tikzpicture}
\draw[help lines, color=gray!30, dashed] 
;
\draw[ gray, thick] (-0.75,0)--(0.75,0) 
;
\draw[decorate,decoration={zigzag,segment length=4pt}](-0.5,-0.5) -- (0.5,-0.5);
\draw[ gray, thick] (0,-1.08)--(0,0.2) 
;
\end{tikzpicture}
& \begin{tikzpicture}
\draw[ gray, thick] (-0.75,0)--(0.75,0);

\draw[ gray, thick] (0,-1.08)--(0,0.2);
\draw[decorate,decoration={zigzag,segment length=4pt}](-0.5,-0.5) -- (0.5,-0.5);
\foreach \Point in {(-0.3,-0.2), (0.3,-0.2)}{
    \node at \Point {\textbullet};
}
\end{tikzpicture}
&
\begin{tikzpicture}
\draw[gray, thick] (-0.75,0)--(0.75,0) 
;
\draw[decorate,decoration={zigzag,segment length=4pt}](-0.5,-0.5) -- (0.5,-0.5);
\draw[gray, thick] (0,-1.08)--(0,0.2) 
;
\foreach \Point in {(-0.0,-0.2)}{
    \node at \Point {\textbullet};
}
\end{tikzpicture}
&

\begin{tikzpicture}
\draw[ gray, thick] (-0.75,0)--(0.75,0);

\draw[ gray, thick] (0,-1.08)--(0,0.2);
\draw[decorate,decoration={zigzag,segment length=4pt}](-0.5,-0.5) -- (0.5,-0.5);

\end{tikzpicture}

&---
&---
         \\
         \hline
         $T,\mu \neq0$ & \begin{tikzpicture}
\draw[ gray, thick] (-0.75,0)--(0.75,0) 
;
\draw[decorate,decoration={zigzag,segment length=4pt}](-0.5,-0.5) -- (0.5,-0.5);
\draw[ gray, thick] (0,-1.08)--(0,0.2) 
;
\foreach \Point in {(-0.0,-0.2),(-0.3,-0.2), (0.3,-0.2)}{
    \node at \Point {\textbullet};
}
\end{tikzpicture}
         & \begin{tikzpicture}
\draw[gray, thick] (-0.75,0)--(0.75,0) 
;
\draw[decorate,decoration={zigzag,segment length=4pt}](-0.5,-0.5) -- (0.5,-0.5);
\draw[gray, thick] (0,-1.08)--(0,0.2) 
;
\foreach \Point in {(-0.0,-0.2)}{
    \node at \Point {\textbullet};
}
\end{tikzpicture}
&

\begin{tikzpicture}
\draw[gray, thick] (-0.75,0)--(0.75,0) 
;
\draw[decorate,decoration={zigzag,segment length=4pt}](-0.5,-0.5) -- (0.5,-0.5);
\draw[gray, thick] (0,-1.08)--(0,0.2) 
;
\foreach \Point in {(-0.3,-0.2), (0.3,-0.2)}{
    \node at \Point {\textbullet};
}
\end{tikzpicture}
&
\begin{tikzpicture}
\draw[ gray, thick] (-0.75,0)--(0.75,0) 
;
\draw[decorate,decoration={zigzag,segment length=4pt}](-0.5,-0.5) -- (0.5,-0.5);
\draw[ gray, thick] (0,-1.08)--(0,0.2) 
;
\foreach \Point in {(0.0,-0.2)}{
    \node at \Point {\textbullet};
}
\end{tikzpicture}
&
\begin{tikzpicture}
\draw[help lines, color=gray!30, dashed] 
;
\draw[ gray, thick] (-0.75,0)--(0.75,0) 
;
\draw[decorate,decoration={zigzag,segment length=4pt}](-0.5,-0.5) -- (0.5,-0.5);
\draw[ gray, thick] (0,-1.08)--(0,0.2) 
;
\end{tikzpicture}
&
\begin{tikzpicture}
\draw[help lines, color=gray!30, dashed] 
;
\draw[ gray, thick] (-0.75,0)--(0.75,0) 
;
\draw[decorate,decoration={zigzag,segment length=4pt}](-0.5,-0.5) -- (0.5,-0.5);
\draw[ gray, thick] (0,-1.08)--(0,0.2) 
;
\foreach \Point in {(-0.3,-0.2), (0.3,-0.2)}{
    \node at \Point {\textbullet};
}
\end{tikzpicture}
&
\begin{tikzpicture}
\draw[ gray, thick] (-0.75,0)--(0.75,0) 
;
\draw[decorate,decoration={zigzag,segment length=4pt}](-0.5,-0.5) -- (0.5,-0.5);
\draw[ gray, thick] (0,-1.08)--(0,0.2) 
;
\foreach \Point in {(-0.,-0.2)}{
    \node at \Point {\textbullet};
}
\end{tikzpicture}

\\

         \hline
         $T,\mu,\Gamma_\parallel \neq 0$
&

\begin{tikzpicture}
\draw[gray, thick] (-0.75,0)--(0.75,0);
\draw[gray, thick] (0,-1.08)--(0,0.2);
\draw[->,red, thick] (0.0,-0.35)--(-0.3,-0.35);
\draw[->,red, thick] (-0.0,-0.35)--(0.3,-0.35);
\draw[->_,red, thick] (-0.0,-0.2)--(0.0,-0.35) ;
\draw[->_,red, thick] (-0.0,-0.6)--(0.0,-0.35) ;
\draw[decorate,decoration={zigzag,segment length=4pt}] (-0.5,-0.7) -- (0.5,-0.7);
\node at (-0.0,-0.5) [circle,black,fill,inner sep=1.5pt]{};
\node at (0.0,-0.2) [circle,black,fill,inner sep=1.5pt]{};
\end{tikzpicture}

&
\begin{tikzpicture}
\draw[help lines, color=gray!30, dashed] 
;
\draw[ gray, thick] (-0.75,0)--(0.75,0) 
;
\draw[decorate,decoration={zigzag,segment length=4pt}] (-0.5,-0.7) -- (0.5,-0.7);
\draw[ gray, thick] (0,-1.08)--(0,0.2) 
;
\foreach \Point in {(-0.0,-0.2)}{
    \node at \Point {\textbullet};
}
\end{tikzpicture}
&

\begin{tikzpicture}
\draw[gray, thick] (-0.75,0)--(0.75,0);
\draw[gray, thick] (0,-1.08)--(0,0.2);
\draw[->,red, thick] (0.0,-0.35)--(-0.3,-0.35);
\draw[->,red, thick] (-0.0,-0.35)--(0.3,-0.35);
\draw[->_,red, thick] (-0.0,-0.2)--(0.0,-0.35) ;
\draw[->_,red, thick] (-0.0,-0.6)--(0.0,-0.35) ;
\draw[decorate,decoration={zigzag,segment length=4pt}] (-0.5,-0.7) -- (0.5,-0.7);
\node at (-0.0,-0.5) [circle,black,fill,inner sep=1.5pt]{};
\node at (0.0,-0.2) [circle,black,fill,inner sep=1.5pt]{};
\end{tikzpicture}

&
\begin{tikzpicture}
\draw[ gray, thick] (-0.75,0)--(0.75,0) 
;
\draw[decorate,decoration={zigzag,segment length=4pt}] (-0.5,-0.7) -- (0.5,-0.7);
\draw[ gray, thick] (0,-1.08)--(0,0.2) 
;
\foreach \Point in {(0.0,-0.2)}{
    \node at \Point {\textbullet};
}
\end{tikzpicture}
&

\begin{tikzpicture}
\draw[ gray, thick] (-0.75,0)--(0.75,0) 
;
\draw[decorate,decoration={zigzag,segment length=4pt}] (-0.5,-0.7) -- (0.5,-0.7);
\draw[ gray, thick] (0,-1.08)--(0,0.2) 
;
\end{tikzpicture}
&
\begin{tikzpicture}
\draw[gray, thick] (-0.75,0)--(0.75,0);
\draw[gray, thick] (0,-1.08)--(0,0.2);
\draw[->,red, thick] (0.0,-0.35)--(-0.3,-0.35);
\draw[->,red, thick] (-0.0,-0.35)--(0.3,-0.35);
\draw[->_,red, thick] (-0.0,-0.2)--(0.0,-0.35) ;
\draw[->_,red, thick] (-0.0,-0.6)--(0.0,-0.35) ;
\draw[decorate,decoration={zigzag,segment length=4pt}] (-0.5,-0.7) -- (0.5,-0.7);
\node at (-0.0,-0.5) [circle,black,fill,inner sep=1.5pt]{};
\node at (0.0,-0.2) [circle,black,fill,inner sep=1.5pt]{};
\end{tikzpicture}

&

\begin{tikzpicture}
\draw[ gray, thick] (-0.75,0)--(0.75,0) 
;
\draw[decorate,decoration={zigzag,segment length=4pt}] (-0.5,-0.7) -- (0.5,-0.7);
\draw[ gray, thick] (0,-1.08)--(0,0.2) 
;
\foreach \Point in {(-0.,-0.2)}{
    \node at \Point {\textbullet};
}
\end{tikzpicture}
\\
\hline

$T,\mu,\Gamma_\perp \neq0$
&
\begin{tikzpicture}
\draw[ gray, thick] (-0.75,0)--(0.75,0) 
;
\draw[decorate,decoration={zigzag,segment length=4pt}](-0.5,-0.5) -- (0.5,-0.5);
\draw[ gray, thick] (0,-1.08)--(0,0.2) 
;
\foreach \Point in {(-0.0,-0.2),(-0.3,-0.2), (0.3,-0.2)}{
    \node at \Point {\textbullet};
}
\end{tikzpicture}
         & \begin{tikzpicture}
\draw[gray, thick] (-0.75,0)--(0.75,0) 
;
\draw[decorate,decoration={zigzag,segment length=4pt}](-0.5,-0.5) -- (0.5,-0.5);
\draw[gray, thick] (0,-1.08)--(0,0.2) 
;
\foreach \Point in {(-0.0,-0.2)}{
    \node at \Point {\textbullet};
}
\end{tikzpicture}
&

\begin{tikzpicture}
\draw[ gray, thick] (-0.75,0)--(0.75,0);

\draw[ gray, thick] (0,-1.08)--(0,0.2);
\draw[decorate,decoration={zigzag,segment length=4pt}](-0.5,-0.5) -- (0.5,-0.5);
\foreach \Point in {(-0.3,-0.2), (0.3,-0.2)}{
    \node at \Point {\textbullet};
}
\end{tikzpicture}

&
\begin{tikzpicture}
\draw[gray, thick] (-0.75,0)--(0.75,0) ;

\draw[gray, thick] (0,-1.08)--(0,0.2);
\draw[decorate,decoration={zigzag,segment length=4pt}](-0.5,-0.5) -- (0.5,-0.5);
\foreach \Point in {(-0.,-0.2)}{
    \node at \Point {\textbullet};
}
\end{tikzpicture}
&
\begin{tikzpicture}
\draw[gray, thick] (-0.75,0)--(0.75,0) ;

\draw[gray, thick] (0,-1.08)--(0,0.2);
\draw[decorate,decoration={zigzag,segment length=4pt}](-0.5,-0.5) -- (0.5,-0.5);

\end{tikzpicture}
&
\begin{tikzpicture}
\draw[ gray, thick] (-0.75,0)--(0.75,0);

\draw[ gray, thick] (0,-1.08)--(0,0.2);
\draw[decorate,decoration={zigzag,segment length=4pt}](-0.5,-0.5) -- (0.5,-0.5);
\foreach \Point in {(-0.3,-0.2), (0.3,-0.2)}{
    \node at \Point {\textbullet};
}
\end{tikzpicture}
&

\begin{tikzpicture}
\draw[gray, thick] (-0.75,0)--(0.75,0) ;

\draw[gray, thick] (0,-1.08)--(0,0.2);
\draw[decorate,decoration={zigzag,segment length=4pt}](-0.5,-0.5) -- (0.5,-0.5);
\foreach \Point in {(-0.,-0.2)}{
    \node at \Point {\textbullet};
}
\end{tikzpicture}
         \\ \hline
    \end{tabular}
    }
    \caption{A `periodic table' of analytic structures of $T^{\mu\nu}$ and $J^\mu$ correlators in $3+1$ dimensions. The first row $T\neq 0$ corresponds to results found in \cite{Romatschke:2015gic}. The second and third rows are a momentum dissipation ($\Gamma_\parallel$ is dissipation in the direction parallel to the perturbations' wavevector, while $\Gamma_\perp$ is dissipation in the transverse direction) extension to the first row. The fourth row $T,\mu\neq 0,$ is the thermoelectric case. 
    The fifth and sixth row are the momentum dissipating thermoelectric case. Hydrodynamic poles are denoted by black points. The logarithmic branch cut is denoted by a squiggly line. In the case of nonzero $\Gamma_\parallel$, we denote the movement of the poles with increasing $k$ with red lines, indicating that there is a collision at $k\sim \Gamma_\parallel$. After the collision, the poles acquire a real part and become propagating. We provide more details of the analytic structure of 
    the spin 0 $\langle JJ\rangle_{T,\mu\neq0}$ in 
    Figure~\ref{fig:00Structure}.
    }\label{table-3plus1}
\end{table}

\begin{table}[h!]
    \centering
    
    \resizebox{\columnwidth}{!}{%
    \begin{tabular}{c|c|c|c|c|c|c|c|c|}
         &
       \multicolumn{2}{c|}{
         $\langle J J \rangle$}
         &
         \multicolumn{2}{c|}{
         $\langle TT \rangle$}
         &
         \multicolumn{2}{c|}{
         $\langle TJ \rangle$}
       \\
       \hline
          & even  & odd
         & even & odd  & even
         &odd
         \\
         \hline
          $T\neq 0$ & 

\begin{tikzpicture}
\draw[ gray,thick] (-0.75,0)--(0.75,0) 
;
\draw[gray, thick] (0,-0.75)--(0,0.2) 
;
\draw[decorate,decoration={zigzag,segment length=4pt}] (0.5,-0.5) arc (0:-180:0.5) ;
\draw[->_,blue, thick] (-0.0,-0.2)--(0.0,-0.5) 
;
\draw[->_,blue, thick] (-0.0,-0.5)--(0.25,-0.5) 
;
\draw[->_,blue, thick] (-0.0,-0.5)--(-0.25,-0.5) 
;
\draw[->_,blue, thick] (-0.0,-0.8)--(0.0,-0.5) 
;

\foreach \Point in {(-0.0,-0.2)}{
    \node at \Point {\textbullet};
}

\node at (-0.0,-0.8) [circle,applegreen,fill,inner sep=1.5pt]{};
\end{tikzpicture}
&
\begin{tikzpicture}
\draw[ gray, thick] (-0.75,0)--(0.75,0) 
;
\draw[decorate,decoration={zigzag,segment length=4pt}](-0.5,-0.5) -- (0.5,-0.5);
\draw[ gray, thick] (0,-1.08)--(0,0.2) 
;
\end{tikzpicture}

&
\begin{tikzpicture}
\draw[ gray, thick] (-0.75,0)--(0.75,0) 
;
\draw[decorate,decoration={zigzag,segment length=4pt}](-0.5,-0.5) -- (0.5,-0.5);
\draw[ gray, thick] (0,-1.08)--(0,0.2) 
;
\foreach \Point in {(-0.3,-0.2), (0.3,-0.2)}{
    \node at \Point {\textbullet};
}
\end{tikzpicture}
&
\begin{tikzpicture}
\draw[ gray, thick] (-0.75,0)--(0.75,0);
\draw[decorate,decoration={zigzag,segment length=4pt}](-0.5,-0.5) -- (0.5,-0.5);
\draw[ gray, thick] (0,-1.08)--(0,0.2) 
;
\foreach \Point in {(-0.,-0.2)}{
    \node at \Point {\textbullet};
}
\end{tikzpicture}

&
---
& 
---

 \\

        \hline
         $T, \Gamma_\parallel\neq0$  & 
\begin{tikzpicture}
\draw[ gray,thick] (-0.75,0)--(0.75,0) 
;
\draw[gray, thick] (0,-0.75)--(0,0.2) 
;
\draw[decorate,decoration={zigzag,segment length=4pt}] (0.5,-0.5) arc (0:-180:0.5) ;
\draw[->_,blue, thick] (-0.0,-0.2)--(0.0,-0.5) 
;
\draw[->_,blue, thick] (-0.0,-0.5)--(0.25,-0.5) 
;
\draw[->_,blue, thick] (-0.0,-0.5)--(-0.25,-0.5) 
;
\draw[->_,blue, thick] (-0.0,-0.8)--(0.0,-0.5) 
;

\foreach \Point in {(-0.0,-0.2)}{
    \node at \Point {\textbullet};
}

\node at (-0.0,-0.8) [circle,applegreen,fill,inner sep=1.5pt]{};
\end{tikzpicture}
&
\begin{tikzpicture}
\draw[ gray, thick] (-0.75,0)--(0.75,0) 
;
\draw[decorate,decoration={zigzag,segment length=4pt}](-0.5,-0.5) -- (0.5,-0.5);
\draw[ gray, thick] (0,-1.08)--(0,0.2) 
;
\end{tikzpicture}

&
\begin{tikzpicture}
\draw[gray, thick] (-0.75,0)--(0.75,0);
\draw[gray, thick] (0,-1.08)--(0,0.2);
\draw[->,red, thick] (0.0,-0.35)--(-0.3,-0.35);
\draw[->,red, thick] (-0.0,-0.35)--(0.3,-0.35);
\draw[->_,red, thick] (-0.0,-0.2)--(0.0,-0.38) ;
\draw[->_,red, thick] (-0.0,-0.5)--(0.0,-0.32) ;
\draw[decorate,decoration={zigzag,segment length=4pt}] (-0.5,-0.7) -- (0.5,-0.7);
\node at (-0.0,-0.5) [circle,black,fill,inner sep=1.5pt]{};
\node at (0.0,-0.2) [circle,black,fill,inner sep=1.5pt]{};
\end{tikzpicture}

&
\begin{tikzpicture}
\draw[gray, thick] (-0.75,0)--(0.75,0) 
;
\draw[decorate,decoration={zigzag,segment length=4pt}] (-0.5,-0.7) -- (0.5,-0.7);
\draw[gray, thick] (0,-1.08)--(0,0.2);
\foreach \Point in {(-0.0,-0.2)}{
    \node at \Point {\textbullet};
}
\end{tikzpicture}
&---
&---
         \\

         \hline
         $T,  \Gamma_\perp\neq0$  & 
\begin{tikzpicture}
\draw[ gray,thick] (-0.75,0)--(0.75,0) 
;
\draw[gray, thick] (0,-0.75)--(0,0.2) 
;
\draw[decorate,decoration={zigzag,segment length=4pt}] (0.5,-0.5) arc (0:-180:0.5) ;
\draw[->_,blue, thick] (-0.0,-0.2)--(0.0,-0.5) 
;
\draw[->_,blue, thick] (-0.0,-0.5)--(0.25,-0.5) 
;
\draw[->_,blue, thick] (-0.0,-0.5)--(-0.25,-0.5) 
;
\draw[->_,blue, thick] (-0.0,-0.8)--(0.0,-0.5) 
;

\foreach \Point in {(-0.0,-0.2)}{
    \node at \Point {\textbullet};
}

\node at (-0.0,-0.8) [circle,applegreen,fill,inner sep=1.5pt]{};
\end{tikzpicture}
&
\begin{tikzpicture}
\draw[ gray, thick] (-0.75,0)--(0.75,0) 
;
\draw[decorate,decoration={zigzag,segment length=4pt}](-0.5,-0.5) -- (0.5,-0.5);
\draw[ gray, thick] (0,-1.08)--(0,0.2) 
;
\end{tikzpicture}

& \begin{tikzpicture}
\draw[ gray, thick] (-0.75,0)--(0.75,0);

\draw[ gray, thick] (0,-1.08)--(0,0.2);
\draw[decorate,decoration={zigzag,segment length=4pt}](-0.5,-0.5) -- (0.5,-0.5);
\foreach \Point in {(-0.3,-0.2), (0.3,-0.2)}{
    \node at \Point {\textbullet};
}
\end{tikzpicture}
&
\begin{tikzpicture}
\draw[gray, thick] (-0.75,0)--(0.75,0) 
;
\draw[decorate,decoration={zigzag,segment length=4pt}](-0.5,-0.5) -- (0.5,-0.5);
\draw[gray, thick] (0,-1.08)--(0,0.2);
\foreach \Point in {(-0.0,-0.2)}{
    \node at \Point {\textbullet};
}
\end{tikzpicture}
&---
&---

         \\

\hline
         $T,\mu \neq0$ & 
\begin{tikzpicture}
\draw[ gray,thick] (-0.75,0)--(0.75,0) 
;
\draw[gray, thick] (0,-0.75)--(0,0.2) 
;
\draw[decorate,decoration={zigzag,segment length=4pt}] (0.5,-0.5) arc (0:-180:0.5) ;
\draw[->_,blue, thick] (-0.0,-0.2)--(0.0,-0.5) 
;
\draw[->_,blue, thick] (-0.0,-0.5)--(0.25,-0.5) 
;
\draw[->_,blue, thick] (-0.0,-0.5)--(-0.25,-0.5) 
;
\draw[->_,blue, thick] (-0.0,-0.8)--(0.0,-0.5) 
;

\foreach \Point in {(-0.3,-0.2), (0.3,-0.2),(-0.0,-0.2)}{
    \node at \Point {\textbullet};
}
\node at (-0.0,-0.8) [circle,applegreen,fill,inner sep=1.5pt]{};
\end{tikzpicture}
&
\begin{tikzpicture}
\draw[ gray, thick] (-0.75,0)--(0.75,0);
\draw[decorate,decoration={zigzag,segment length=4pt}](-0.5,-0.5) -- (0.5,-0.5);
\draw[ gray, thick] (0,-1.08)--(0,0.2);
\foreach \Point in {(-0.,-0.2)}{
    \node at \Point {\textbullet};
}
\end{tikzpicture}
&
\begin{tikzpicture}
\draw[ gray, thick] (-0.75,0)--(0.75,0) 
;
\draw[decorate,decoration={zigzag,segment length=4pt}](-0.5,-0.5) -- (0.5,-0.5);
\draw[ gray, thick] (0,-1.08)--(0,0.2) 
;
\foreach \Point in {(-0.3,-0.2), (0.3,-0.2)}{
    \node at \Point {\textbullet};
}
\end{tikzpicture}
&
\begin{tikzpicture}
\draw[ gray, thick] (-0.75,0)--(0.75,0) 
;
\draw[decorate,decoration={zigzag,segment length=4pt}](-0.5,-0.5) -- (0.5,-0.5);
\draw[ gray, thick] (0,-1.08)--(0,0.2) 
;
\foreach \Point in {(-0.,-0.2)}{
    \node at \Point {\textbullet};
}
\end{tikzpicture}
&

\begin{tikzpicture}
\draw[ gray, thick] (-0.75,0)--(0.75,0);
\draw[decorate,decoration={zigzag,segment length=4pt}](-0.5,-0.5) -- (0.5,-0.5);
\draw[ gray, thick] (0,-1.08)--(0,0.2);
\foreach \Point in {(-0.3,-0.2), (0.3,-0.2)}{
    \node at \Point {\textbullet};
}
\end{tikzpicture}

&
\begin{tikzpicture}
\draw[ gray, thick] (-0.75,0)--(0.75,0);
\draw[decorate,decoration={zigzag,segment length=4pt}](-0.5,-0.5) -- (0.5,-0.5);
\draw[ gray, thick] (0,-1.08)--(0,0.2);
\foreach \Point in {(-0.,-0.2)}{
    \node at \Point {\textbullet};
}
\end{tikzpicture}

\\
\hline
$T,\mu,\Gamma_\parallel\neq0$

&

\begin{tikzpicture}

\draw[gray, thick] (-0.75,0)--(0.75,0);
\draw[gray, thick] (0,-0.9)--(0,0.2);
\draw[blue, thick] (-0.0,-0.35)--(0.0,-0.5) 
;
\draw[->,blue, thick] (-0.0,-0.73)--(0.3,-0.73) 
;
\draw[->,blue, thick] (-0.0,-0.73)--(-0.3,-0.73) 
;
\draw[->_,blue, thick] (-0.0,-0.58)--(0.0,-0.77) 
;
\draw[->_,blue, thick] (-0.0,-0.88)--(0.0,-0.70) 
;
\draw[->,red, thick] (0.0,-0.27)--(-0.3,-0.27);
\draw[->,red, thick] (-0.0,-0.27)--(0.3,-0.27);
\draw[->_,red, thick] (-0.0,-0.1)--(0.0,-0.3) ;
\draw[->_,red, thick] (-0.0,-0.4)--(0.0,-0.24) ;
\draw[red, thick] (-0.0,-0.45)--(0.0,-0.35) ;

\draw[decorate,decoration={zigzag,segment length=4pt}] (0.5,-0.5) arc (-0:-180:0.5) ;
\node at (-0.0,-0.86) [circle,applegreen,fill,inner sep=1.2pt]{};
\node at (0.0,-0.1) [circle,black,fill,inner sep=1.2pt]{};
\node at (0.0,-0.43) [circle,black,fill,inner sep=1.2pt]{};
\node at (0.0,-0.58) [circle,black,fill,inner sep=1.2pt]{};
\end{tikzpicture}

&
\begin{tikzpicture}
\draw[ gray, thick] (-0.75,0)--(0.75,0) ;
\draw[decorate,decoration={zigzag,segment length=4pt}] (-0.5,-0.7) -- (0.5,-0.7);
\draw[ gray, thick] (0,-1.08)--(0,0.2);
\foreach \Point in {(-0.,-0.2)}{
    \node at \Point {\textbullet};
}
\end{tikzpicture}
&

\begin{tikzpicture}
\draw[gray, thick] (-0.75,0)--(0.75,0);
\draw[gray, thick] (0,-1.08)--(0,0.2);
\draw[->,red, thick] (0.0,-0.35)--(-0.3,-0.35);
\draw[->,red, thick] (-0.0,-0.35)--(0.3,-0.35);
\draw[->_,red, thick] (-0.0,-0.2)--(0.0,-0.38) ;
\draw[->_,red, thick] (-0.0,-0.5)--(0.0,-0.32) ;
\draw[decorate,decoration={zigzag,segment length=4pt}] (-0.5,-0.7) -- (0.5,-0.7);
\node at (-0.0,-0.5) [circle,black,fill,inner sep=1.5pt]{};
\node at (0.0,-0.2) [circle,black,fill,inner sep=1.5pt]{};
\end{tikzpicture}

&

\begin{tikzpicture}
\draw[gray, thick] (-0.75,0)--(0.75,0) ;

\draw[gray, thick] (0,-1.08)--(0,0.2);
\draw[decorate,decoration={zigzag,segment length=4pt}] (-0.5,-0.7) -- (0.5,-0.7);
\foreach \Point in {(-0.,-0.2)}{
    \node at \Point {\textbullet};
}
\end{tikzpicture}

&
\begin{tikzpicture}
\draw[gray, thick] (-0.75,0)--(0.75,0);
\draw[gray, thick] (0,-1.08)--(0,0.2);
\draw[->,red, thick] (0.0,-0.35)--(-0.3,-0.35);
\draw[->,red, thick] (-0.0,-0.35)--(0.3,-0.35);
\draw[->_,red, thick] (-0.0,-0.2)--(0.0,-0.38) ;
\draw[->_,red, thick] (-0.0,-0.5)--(0.0,-0.32) ;
\draw[decorate,decoration={zigzag,segment length=4pt}] (-0.5,-0.7) -- (0.5,-0.7);
\node at (-0.0,-0.5) [circle,black,fill,inner sep=1.5pt]{};
\node at (0.0,-0.2) [circle,black,fill,inner sep=1.5pt]{};
\end{tikzpicture}

&

\begin{tikzpicture}
\draw[ gray, thick] (-0.75,0)--(0.75,0);
\draw[decorate,decoration={zigzag,segment length=4pt}] (-0.5,-0.7) -- (0.5,-0.7);
\draw[ gray, thick] (0,-1.08)--(0,0.2);
\foreach \Point in {(-0.,-0.2)}{
    \node at \Point {\textbullet};
}
\end{tikzpicture}

         \\
         \hline
         $T,\mu,\Gamma_\perp \neq 0$
&
\begin{tikzpicture}
\draw[ gray,thick] (-0.75,0)--(0.75,0) 
;
\draw[gray, thick] (0,-0.75)--(0,0.2) 
;
\draw[decorate,decoration={zigzag,segment length=4pt}] (0.5,-0.5) arc (0:-180:0.5) ;
\draw[->_,blue, thick] (-0.0,-0.2)--(0.0,-0.5) 
;
\draw[->_,blue, thick] (-0.0,-0.5)--(0.25,-0.5) 
;
\draw[->_,blue, thick] (-0.0,-0.5)--(-0.25,-0.5) 
;
\draw[->_,blue, thick] (-0.0,-0.8)--(0.0,-0.5) 
;

\foreach \Point in {(-0.3,-0.2), (0.3,-0.2),(-0.0,-0.2)}{
    \node at \Point {\textbullet};
}
\node at (-0.0,-0.8) [circle,applegreen,fill,inner sep=1.5pt]{};
\end{tikzpicture}
&
\begin{tikzpicture}
\draw[ gray, thick] (-0.75,0)--(0.75,0);
\draw[decorate,decoration={zigzag,segment length=4pt}](-0.5,-0.5) -- (0.5,-0.5);
\draw[ gray, thick] (0,-1.08)--(0,0.2);
\foreach \Point in {(-0.,-0.2)}{
    \node at \Point {\textbullet};
}
\end{tikzpicture}
&
\begin{tikzpicture}
\draw[ gray, thick] (-0.75,0)--(0.75,0);
\draw[decorate,decoration={zigzag,segment length=4pt}](-0.5,-0.5) -- (0.5,-0.5);
\draw[ gray, thick] (0,-1.08)--(0,0.2);
\foreach \Point in {(-0.3,-0.2), (0.3,-0.2)}{
    \node at \Point {\textbullet};
}
\end{tikzpicture}
&
\begin{tikzpicture}
\draw[ gray, thick] (-0.75,0)--(0.75,0);
\draw[decorate,decoration={zigzag,segment length=4pt}](-0.5,-0.5) -- (0.5,-0.5);
\draw[ gray, thick] (0,-1.08)--(0,0.2);
\foreach \Point in {(-0.,-0.2)}{
    \node at \Point {\textbullet};
}
\end{tikzpicture}

&
\begin{tikzpicture}
\draw[ gray, thick] (-0.75,0)--(0.75,0);
\draw[decorate,decoration={zigzag,segment length=4pt}](-0.5,-0.5) -- (0.5,-0.5);
\draw[ gray, thick] (0,-1.08)--(0,0.2);
\foreach \Point in {(-0.3,-0.2), (0.3,-0.2)}{
    \node at \Point {\textbullet};
}
\end{tikzpicture}
&

\begin{tikzpicture}
\draw[ gray, thick] (-0.75,0)--(0.75,0) 
;
\draw[decorate,decoration={zigzag,segment length=4pt}](-0.5,-0.5) -- (0.5,-0.5);
\draw[ gray, thick] (0,-1.08)--(0,0.2) 
;
\foreach \Point in {(-0.,-0.2)}{
    \node at \Point {\textbullet};
}
\end{tikzpicture}

\\
\hline

    \end{tabular}
    }

    \caption{A `periodic table' of analytic structures of $T^{\mu\nu}$ and $J^\mu$ correlators in $2+1$ dimensions. In addition to the labelling used in Table~\ref{table-3plus1}, the new structure of colliding poles at $\tau_R k\sim1$ (where $\tau_R$ is the relaxation time)
    is depicted with blue arrows. The green dots are the special gapped modes in $2+1$ dimensions that remain gapped when $\Gamma = 0$. For the even channel $\langle JJ \rangle$ correlators, the branch cut is chosen and drawn (see Appendix~\ref{app:ComplexAnalysis}) so as to make the representation of the gapped pole (in green) clearer. We note that the shape of the cut can indeed be arbitrarily chosen so long as it connects the two branch points. We include more details of the even $\langle JJ\rangle _{T\neq0}$ in Figure~\ref{fig:poles2plus1JJ}, the even $\langle TT\rangle_{T,\Gamma_\parallel\neq0}$ in 
    Figure~\ref{fig:2dTT-mombr} and
    the even $\langle JJ\rangle _{T, \mu,\Gamma_\parallel\neq0}$ in Figure~\ref{fig:jj-TmuG}.
    }\label{table-2plus1}
\end{table}

{The outline of the paper is as follows: in Section~\ref{sec:setup}, we first discuss the RTA kinetic theory. We present the method for the calculation of correlation functions and discuss the relevant aspects of the thermoelectric effect. With the basics established, we state the complete set of analytic results for the correlators in Section~\ref{sec:correlators}, which represents the main result of the present work.  We then analyze both in $2+1$ and in $3+1$ dimensions various details of the uncharged correlators in Section~\ref{sec:charge-neutral}, the uncharged correlators with momentum breaking in Section~\ref{sec:uncharged-mombr}, the charged case in Section~\ref{sec:charged} and the charged momentum breaking case in Section~\ref{sec:charged-mombr}. Conclusions and future directions are then discussed in Section~\ref{sec:Conclusion}. Appendix~\ref{app:ComplexAnalysis} is devoted to certain relevant details of complex analysis that are necessary for understanding physical properties of computed spectra. Then, to facilitate a stringent consistency check of our kinetic theory results in the (hydrodynamic) low-energy limit, we also perform an independent hydrodynamic calculation of all studied correlators in Appendix~\ref{app:canonicalapproach}. Finally, Appendix~\ref{app:contact} contains details of the kinetic theory calculation that pertain to contact terms.}

\section{Set up}\label{sec:setup}

\subsection{Kinetic theory in the RTA}

In this section, we lay down the basic ideas of RTA kinetic theory, following the conventions of \cite{Romatschke:2015gic}. The Boltzmann equation for the on-shell one particle distribution function, $f(t,\textbf{x},\textbf{p})$, is
\begin{align}\label{eq:Boltzmann}
    \left[p^\mu \partial_\mu+ F^\mu \nabla_{p^\mu}\right]f=C[f],
\end{align}
where $F^\mu$ represents the forces acting on a particle and $C[f]$ is the collision kernel. The electromagnetic and gravitational forces are 
\begin{align}
    F^\mu&=F^{\mn}p_\nu,\\
    F^\mu&=-\Gamma^\mu_{\ab} p^\alpha p^\beta,
\end{align}
respectively. $F^\mn$ is the electromagnetic field strength tensor and $\Gamma^\mu_{\ab}$ is the Christoffel symbol. Here, we will take the collision kernel to be the Anderson-Witting \cite{anderson} RTA kernel, namely
\begin{align}\label{collision}
    C[f]=\frac{p^\mu u_\mu}{\tau_R}\left(f-f^{\rm eq}\right),
\end{align}
where 
\begin{align}\label{eq:equilibrium-f}
f^{\rm eq}(t,\textbf{x},\textbf{p})=e^{(p\cdot u(t,\textbf{x})+\mu(t,\textbf{x}))/T(t,\textbf{x})},
\end{align}
is the equilibrium distribution function to which the system relaxes on the timescale of $\tau_R$. Furthermore, the lowest moments of the distribution function provide the current and energy momentum tensor in $d+1$ dimensions
\begin{align}\label{current}
    J^\mu &= \int \frac{d^d p}{(2\pi)^d}\frac{p^\mu}{p^0} f,\\
    T^\mn&=  \int \frac{d^d p}{(2\pi)^d}\frac{p^\mu p^\nu}{p^0} f \label{emt},
\end{align}
which are conserved: 
\begin{align}
   \nabla_\mu T^\mn=0, \quad \nabla_\mu J^\mu = 0.
\end{align}
For conservation to hold, the RTA prescription requires moments of \eqref{collision} to vanish, which results in the matching conditions
\begin{align}\label{matching}
u_\nu \int \frac{d^dp}{(2\pi)^d p^0} p^\mu p^\nu\, (f-f^{eq})&=0, \quad
u_\nu \int \frac{d^dp}{(2\pi)^d p^0}  p^\nu\, (f-f^{eq})=0.
\end{align}

Furthermore, it will also be convenient to define the following thermodynamic quantities in equilibrium for massless particles using the Maxwell-Boltzmann distribution function:
\begin{align}\label{eq:EoS1}
n_0 &= \frac{2 \pi^{d/2}}{(2\pi)^d}\frac{\Gamma(d)}{\Gamma(d/2)}e^{\frac{\mu_0}{T_0}} T_0^d,\\
    \varepsilon_0 &= d\, T_0 n_0,\quad P_0 = \frac{1}{d} \varepsilon_0\label{eq:EoS},\\
    \chi &= \int\frac{d^d p}{(2\pi)^d } \frac{f_0(p)}{T_0}\,\Omega_d \,p^{d-1} = \frac{n_0}{T_0},\label{eq:chi}\\
    d(\varepsilon_0+P_0) &= \int \frac{d^d p}{(2\pi)^d }\frac{f_0(p)}{T_0}\,\Omega_d\, p^{d+1}\label{eq:enth} ,
\end{align}
where $n_0$ is the equilibrium number density, $\varepsilon_0$ is the equilibrium energy density, $P_0$ is the equilibrium pressure, $\chi$ is the static susceptibility, $p = |\textbf{p}|$, $f_0(p)=e^{-(p-\mu_0)/T_0}$ 
and $\Omega_d = 2 \pi^{d/2}/\Gamma(d/2)$ is the solid angle in $d$ spatial dimensions. The two relevant cases for our discussion are $d=2$ and $d=3$ where $\Omega_2 = 2\pi$ and $\Omega_3 = 4\pi$, respectively.

\subsection{Computation of the correlators} \label{sec:thermo-rta}

We turn on the external sources, $\delta A_\mu$ and $\delta g_{\mn}$,  inducing a change in the temperature, chemical potential and the four velocity
\begin{align}
    T(t,\textbf{x})&= T_0 + \delta T (t,\textbf{x}),\\
    \mu(t,\textbf{x})&= \mu_0 + \delta \mu (t,\textbf{x}),\\
    u^\mu (t,\textbf{x})&=(1,\textbf{0})+ \delta u^\mu (t,\textbf{x}).
\end{align}
Note that we normalize $u^\mu u_\mu=-1.$ The external fields, by causing a change in the macroscopic variables, modify the equilibrium distribution function and the distribution function
\begin{align}
    f^{\rm eq}(t,\textbf{x},\textbf{p})&= f_0(\textbf{p})+\delta f^{\rm eq}(t,\textbf{x},\textbf{p}),\\
f(t,\textbf{x},\textbf{p})&=f_0(\textbf{p})+\delta f(t,\textbf{x},\textbf{p}).
\end{align}
For the purposes of this paper, we will expand around the Maxwell-Boltzmann distribution $f_0(\textbf{p}) = e^{-(p^0-\mu_0)/T_0}$. We will also be working with massless particles, $p^2=0.$

We see that the change in the equilibrium distribution \eqref{eq:equilibrium-f} is
\begin{equation}\label{eq:eq-change}
    \delta f^{\rm eq} = \frac{f_0}{T_0}\left( \delta \mu + p^0 \textbf{v}\cdot\delta\textbf{u} + (p^0-\mu_0)\frac{\delta T}{T_0}\right),
\end{equation}
where $\textbf{v}=\textbf{p}/p^0.$

In Fourier space\footnote{The convention we follow for Fourier transforms is $f(\omega,\textbf{k})=\int_{-\infty}^\infty dt \int d^3 x \, e^{i \omega t - i \textbf{k}\cdot \textbf{x}}f(t,\textbf{k})$.}, the linearized Boltzmann equation reads\footnote{Note that we used the identity presented in \cite{Romatschke:2015gic}; $\Gamma^\mu_{\alpha\beta}p_\mu p^\alpha p^\beta = p^\mu \partial_\mu p^2 = 0$ to substitute $\Gamma^i_{\alpha\beta} p_i p^\alpha p^\beta = -\Gamma^0_{\alpha\beta} p_0 p^\alpha p^\beta$.}
\begin{equation}
    \left(-i\omega + i \textbf{k}\cdot\textbf{v}\right)\delta f
    -\frac{f_0}{T_0} \left( \textbf{v}\cdot \textbf{E} - \Gamma^0_{\alpha \beta} p^0 v^\alpha v^\beta\right)  = -\frac{\delta f- \delta f^{\rm eq}}{\tau_R},
\end{equation}
where the electric field is $E^i=\nabla^i A_0-\partial_t A^i$.
The solution is given by 
\begin{equation}\label{bolt-sol}
    \delta f =  \frac{\frac{f_0}{T_0} \tau_R \textbf{E}\cdot\textbf{v} - \frac{f_0}{T_0}\tau_R \Gamma^0_{\alpha \beta} p^0 v^\alpha v^\beta  
    + \delta f_{\text{eq}}}{1 + \tau_R (-i\omega + i \textbf{k}\cdot\textbf{v})}.
\end{equation}

We then compute the change to current and energy momentum tensor via \eqref{current} and \eqref{emt}, respectively. We determine $(\delta \mu, \delta T, \delta \textbf{u})$ by self-consistently solving the matching conditions \eqref{matching} in terms of the external gauge field, $\delta A_\mu,$ and metric perturbations, $\delta g_\mn.$ In other words, we need to self-consistently solve for $(\delta \mu, \delta T, \delta u^i)$ via
\begin{align}
    \delta J^0 &= \delta n(\mu,T),\\
    \delta T^{00} &= \delta \varepsilon(\mu, T),\\
    \delta T^{0i} &=(\varepsilon_0+P_0) \delta u^i.
\end{align}
where $(\delta T, \delta \mu)$ are related to $(\delta n, \delta \varepsilon)$ via
\cite{Kovtun:2012rj}
\begin{align}
    \delta n &=  \frac{\partial n_0}{\partial \mu_0} \delta \mu + \frac{\partial n_0}{\partial T_0} \delta T,\\
    \delta \varepsilon &= \frac{\partial \varepsilon_0}{\partial T_0}\delta T +\frac{\partial \varepsilon_0}{\partial \mu_0}\delta \mu.
\end{align}

We would like to take a moment to highlight that although we are working in the linear regime at the level of perturbations, the non-trivial matching conditions include non-linearities.
Finally, we compute the retarded correlators via the variational principle, namely, by writing 
\begin{align}
    J^\mu&=J^\mu_0-G^{\mu,\nu}_{JJ}\delta A_\nu-\frac{1}{2}G^{\mu,\ab}_{JT}\delta g_{\ab}+\ldots,\\
    T^\mn&=T^\mn_0-\frac{1}{2}G^{\mn,\ab}_{TT}\delta g_{\ab}-G^{\mn,\alpha}_{TJ}\delta A_{\alpha}+\ldots ,
\end{align}
where
\begin{align}\label{correlators}
G_{JJ}^{\mu,\nu}&= -\frac{\delta J^\mu}{\delta A_\nu}, \quad
G_{TJ}^{\mn,\alpha}= -\frac{\delta T^\mn}{\delta A_\alpha}, \quad G_{JT}^{\mu,\alpha\beta} = - 2 \frac{\delta J^\mu}{\delta g_{\ab}}, \quad
   G_{TT}^{\mn,\ab}= -2\frac{\delta T^\mn}{\delta g_\ab}.
\end{align}

\subsection{Momentum breaking}\label{sec:mom-br}

{In theories that study transport (hydrodynamics) with momentum relaxation,
the usual equations of motion (the conservation laws) can be modified to include explicit momentum non-conservation as in equation~\eqref{eq:MomentumBreak},
while leaving the energy conservation intact (see Ref.~\cite{Hartnoll:2007ih,Davison:2013jba,Davison:2014lua}). This in turn modifies the two-point function Ward identities, which are now of the form}
\begin{align}\label{eq:Wards}
    &ik_\mu G^{\mu,\nu}_{JJ} =0 \\
    &i k_\mu G^{\mu,\ab}_{JT} = 0,\\
    & i k_\mu G^{\mn,\alpha}_{TJ}=\begin{cases}
0, & \nu\neq i,\\
 - \Gamma G^{0i,\alpha}_{TJ}, & \nu=i,
\end{cases} \\
    &i k_\mu G^{\mn,\rho\sigma}_{TT}
    = \begin{cases}
0, & \nu\neq i,\\
- \Gamma G^{0i,\rho\sigma}_{TT}, & \nu=i,
\end{cases}\label{wards4}
\end{align}
up to contact terms, which are thoroughly considered in Appendix \ref{app:contact}.

To include momentum relaxation in RTA kinetic theory, 
we modify the collision kernel directly via 
\begin{align}
    C[f]=\frac{p^\mu u_\mu}{\tau_R}\left(f-f_{\rm eq}\right)- \frac{p^\alpha u_\alpha}{T}  \, u_\alpha g^{\alpha\mu} \Gamma_\mn p^\nu f ,
\end{align}
which has the interpretation of the addition of inelastic scattering to the usual RTA. We choose $\Gamma_\mn=\text{diag}(0,\Gamma_\parallel,\Gamma_\perp,\Gamma_\perp)$ in $3+1$ dimensions and $\Gamma_\mn=\text{diag}(0,\Gamma_\parallel,\Gamma_\perp)$ in $2+1$ dimensions.\footnote{It is important to note that $\Gamma_\mn$ is \textit{not} a tensor and hence does not transform covariantly. This stems from the fact that it parametrizes a term explicitly breaking translational symmetry.} Perturbing to linear order, we see that in addition to the terms in \eqref{eq:eq-change}, we have terms proportional to $\Gamma$:
\begin{align}
    \delta f_{\rm eq}^\Gamma = \delta f_{\rm eq} -  \tau_R\, p^i \Gamma_{ij} \left(\delta u^j - \delta g_{0k}\delta^{jk}\right)\frac{f_0}{T_0}.
\end{align}
Note that with this choice of $\Gamma_\mn$ orthogonal to $u^\mu$ in the local rest frame, the leading-order solution is not modified. One can easily see that this choice reproduces our desired form of momentum dissipation in \eqref{eq:MomentumBreak}.

\subsection{Thermoelectric effect} \label{sec:thermo}

We continue with the calculation of the thermoelectric transport matrix in linearized theory:
\begin{equation}\label{eq:TE_TransportMatrix}
    \begin{pmatrix}
        \delta J^i \\
        \delta Q^i
    \end{pmatrix} = \begin{pmatrix}
        \sigma^{ij} & T_0 \alpha^{ij} \\
        T_0 \tilde{\alpha}^{ij} & T_0 \bar{\kappa}^{ij}
    \end{pmatrix}
    \begin{pmatrix}
        E_j \\
        -\frac{1}{T_0}\nabla_j \delta T
    \end{pmatrix},
\end{equation}
where the current and energy momentum tensor is given in \eqref{current} and \eqref{emt}, respectively, and the heat current $\delta Q^i=\delta T^{0i}-\mu_0 \delta J^i$. Since we consider time-reversal invariant systems (i.e. in absence of magnetic effects), it follows that $\alpha = \tilde{\alpha}$ due to Onsager reciprocal relations \cite{onsager}. {This will provide a non-trivial check on the consistency of our results.}

We briefly recall the relations between external perturbations of the metric $\delta g_{\mu\nu}$ and gauge field $\delta A_\mu$ and the sources $\textbf{E}$ and $\nabla T$ of the system's response. For this purpose we summarize the argument of \cite{Hartnoll:2009sz,Herzog:2009xv,Hartnoll:2016apf} in terms more suitable to our setup. We start in Euclidean signature and introduce the temperature rescaled dimensionless time $\bar{t} \equiv t T$, which transforms the $tt$-component of our metric. Consequently, a small change in temperature is equivalent to perturbation in the $\bar t \bar{t}$-component of the metric:
\begin{equation}
    g_{\bar{t}\bar{t}} = \frac{1}{T_0^2} - 2 \frac{\delta T}{T_0^3} \equiv  \eta_{\bar{t}\bar{t}} + \delta g_{\bar{t}\bar{t}},
\end{equation}
where we denote $T_0$ the equilibrium value of temperature. Since we are working in the linearized regime, we can freely deform our perturbations with infinitesimal diffeomorphisms generated by a vector field $\xi_\mu$:
\begin{equation}\label{diffeo}
    \delta g'_{\mu\nu} = \delta g_{\mu\nu} + \partial_\mu \xi_\nu + \partial_\nu \xi_\mu.
\end{equation}
The external gauge field $A_\mu = \bar A_\mu + \delta A_\mu$, where $\bar A_t = \mu_0$, is also transformed via the Lie derivative
\begin{equation}
    A'_\mu = A_\mu + A_\nu \partial_\mu \xi^\nu + \xi^\nu \partial_\nu A_\mu.
\end{equation}
Let us choose
\begin{equation}\label{xi-thermo}
     \xi_\mu = -\frac{1}{i\bar{\omega}} \frac{\delta T}{T_0^3} \delta^{\bar t}_{~\mu} ,
\end{equation}
with the time dependence of $e^{-i\bar{\omega}\bar{t}}$, where $\bar{\omega}$ is the rescaled frequency $\bar{\omega}= \omega/T_0$. One can then show that in this new gauge the temperature perturbation is encoded in the off-diagonal components of the metric perturbation:
\begin{align}
&\delta g'_{\bar{t}\bar{t}} = 0, \quad
\delta g'_{j\bar{t}} =  -\frac{\partial_j \delta T }{ i\bar{\omega} T_0^3}.
\end{align}
In our new gauge the spatial components of the gauge field perturbation take the form of
\begin{equation}
    \delta A'_{\bar{t}}= \delta A_{\bar{t}} -  \frac{\bar A_{\bar{t}}}{T_0}\frac{\delta T}{T_0},\quad\delta A'_j = \delta A_j + \bar A_{\bar{t}} \frac{1}{i\bar{\omega}} \frac{\partial_j \delta T}{T_0^2}.
\end{equation}
The electric field $E_j$ is expressed in the zero momentum limit $k=0$ as simply $E_j = i\omega \delta A_j$. Utilizing this, rescaling back and continuing to Lorentz signature, we obtain the relations
\begin{align}\label{eq:perturbations}
    \delta g'_{tt} = 0, \quad \delta g'_{tj} = -\frac{1}{i\omega}\frac{\partial_j \delta T}{T_0}, \quad \delta A'_j = \frac{E_j}{i \omega} - \mu_0 \frac{1}{i\omega} \frac{\partial_j \delta T}{T_0}.
\end{align}
Expressing $\textbf{E}$ and $\nabla \delta T$ with $\delta A'_\mu$ and $\delta g'_{\mu\nu}$ we write down the thermoelectric transport matrix from \eqref{eq:TE_TransportMatrix} as
\begin{equation}\label{eq:TE_TransportMatrix2}
    \begin{pmatrix}
        \delta J^i \\
        \delta Q^i
    \end{pmatrix} = \begin{pmatrix}
        \sigma^{ij} & \alpha^{ij} T_0 \\
        \tilde{\alpha}^{ij} T_0 & \bar{\kappa}^{ij}  T_0
    \end{pmatrix}
    \begin{pmatrix}
        i\omega \delta A_j' + i\omega \mu_0 \delta g_{tj}' \\
        i\omega \delta g_{tj}'
    \end{pmatrix}.
\end{equation}
In order to obtain the correct behavior of the imaginary components of the conductivities, we must substract the $\omega=0$ part of the appropriate correlator \cite{Herzog:2009xv,hartnoll2018holographic,Davison:2014lua}. With this in mind, we can read off the transport coefficients from \eqref{eq:TE_TransportMatrix2}:
\begin{align}\label{kubo}
    \sigma^{ij}(\omega) &= -\frac{1}{i\omega}\lim_{k\rightarrow 0}\left(G^{ij}_{JJ}(\omega,k) - G^{ij}_{JJ}(0,k)\right),\\
    \tilde{\alpha}^{ij}(\omega) &=  -\frac{1}{i\omega T_0}\lim_{k\rightarrow0}\left(G^{ij}_{QJ}(\omega,k)-G^{ij}_{QJ}(0,k)\right),\label{kubo_alphat}\\
    \alpha^{ij}(\omega) &= -\frac{1}{i\omega T_0}\lim_{k\rightarrow0}\left(G^{ij}_{JQ}(\omega,k)-G^{ij}_{JQ}(0,k)\right),\label{kubo_alpha}\\
    \bar{\kappa}^{ij}(\omega) &= -\frac{1}{i\omega T_0}\lim_{k\rightarrow0}\left(G^{ij}_{QQ}(\omega,k) - G^{ij}_{QQ}(0,k)\right)\label{kubo_kappa},
\end{align}
where the correlators are defined in \eqref{correlators} and
\begin{equation}
     \frac{\delta X^i}{\delta Q^j} = 2\frac{\delta X^i}{\delta g_{tj}} - \mu_0 \frac{\delta X^i}{\delta A_j}.
\end{equation}

\section{Analytic expressions of the correlators}\label{sec:correlators}

The computation of the correlators, as outlined in Section~\ref{sec:thermo-rta}, requires integrating the distribution function \eqref{bolt-sol} over the phase space and solving self-consistently for the relevant macroscopic variables $(\delta n, \delta T, \delta u^i)$. Here we present the main result of the paper, namely, the explicit analytic expressions of all correlators. We will subsequently analyze their structure in the following sections. Without loss of generality, we will align our perturbation to be in the $1-$direction in all cases below.

We note that the angular part of integration over momentum $p$ of the perturbed distribution function $\delta f$ provides the characteristic logarithmic term in $3+1$ dimensions and a square root term in $2+1$ dimensions. The details on the chosen branch cut and the Riemann sheet of these multivalued maps are presented in Appendix \ref{app:ComplexAnalysis}.

We also note that the correlators exhibit certain symmetries in their indices (e.g., the pairwise symmetry $G^{\ab,\mn}=G^{\mn,\ab}$) and satisfy the Ward identities \eqref{eq:Wards}--\eqref{wards4}. As such, we write down the minimum set necessary to reconstruct all of the (nonzero) correlators.

\subsection{Finite temperature correlators}

For completeness, we begin by writing down the correlators from \cite{Romatschke:2015gic}. Turning on only the external gauge field (thereby generating a change in the number density, $\delta n$), we find the following correlators in $3+1$ dimensions
\begin{align}
    G_{JJ}^{0,0}&=-\frac{\chi  (2 k \tau_R+L (\tau_R \omega +i))}{2 k \tau_R+i L},\label{cor-3d-rom00}\\
    G_{JJ}^{2,2}&=-\frac{\chi  \omega  \left(L \left(k^2 \tau_R^2-(\tau_R \omega +i)^2\right)-2 k \tau_R (\tau_R \omega
   +i)\right)}{4 k^3 \tau_R^2}\label{cor-3d-rom22},
\end{align}
where we defined $L\equiv\ln\left(\frac{\omega - k + i/\tau_R}{\omega + k + i/\tau_R}\right)$. Staying in $3+1$ dimensions, we now turn on only the metric perturbation, which leads to 
\begin{align}\label{cor-3d-rom0000}
    G^{00,00}_{TT}&=-3(\varepsilon_0+P_0)\left(1+\frac{ k^2 \tau_R  (2 k \tau_R+L (\tau_R \omega +i))}{2 k^3 \tau_R^2+i
   k^2 L \tau_R+6 i k \tau_R \omega +3 i L \omega  (\tau_R \omega +i)}\right),\\
   G^{22,00}_{TT} &= -3(\varepsilon_0+P_0)\left(\frac{2}{3}+\frac{ \omega  \left(L \left(k^2 \tau_R^2-\tau_R^2 \omega ^2-4 i \tau_R \omega
   +3\right)-2 k \tau_R (\tau_R \omega +3 i)\right)}{4 k^3 \tau_R^2+2 i k^2 L \tau_R+12 i k \tau_R \omega +6 i L \omega  (\tau_R \omega +i)}\right),\\
   G^{02,02}_{TT}&= -\frac{(\varepsilon_0+P_0) }{3 i L \left(k^2 \tau_R^2-(\tau_R \omega +i)^2\right)+2 k \tau_R \left(2 k^2 \tau_R^2-3 i \tau_R \omega +3\right)}\notag\\
   &\times\bigg(3 L (\tau_R \omega +i) \left(k^2 \tau_R^2-(\tau_R \omega
   +i)^2\right)+2 k \tau_R \left(2 k^2 \tau_R^2-3 (\tau_R \omega +i)^2\right)\bigg),\\
   G^{23,23}_{TT}& = -\frac{\omega  (\varepsilon_0 +P_0)}{16 k^5 \tau_R^4}\notag\\
   &\times \left(3 L \left(k^2 \tau_R^2+(1-i \tau_R \omega )^2\right)^2-10 k^3
   \tau_R^3 (\tau_R \omega +i)+6 k \tau_R (\tau_R \omega +i)^3\right).\label{spin-2-thermal}
\end{align}

Next, we provide the same correlators as above, but in $2+1$ dimensions. The $\langle JJ\rangle$ correlators are given by
\begin{align}\label{cor-2d-uncoup00}
   G^{0,0}_{JJ} &= -\frac{\chi  \left(R+i\omega \tau_R  -1\right)}{-1+R},\\
    G^{2,2}_{JJ}&=-\chi \frac{i  \omega \tau_R \left( R+i 
   \omega \tau_R  -1\right)}{k^2 \tau_R^2},
\end{align}
where we introduced the shorthand notation $R \equiv  \sqrt{k^2\tau_R^2 - (\omega \tau_R + i )^2}$. The $\langle TT\rangle$ correlators are 
\begin{align}\label{cor-2d-rom0000}
    G^{00,00}_{TT} &=-2 (\varepsilon_0+P_0)\left(1+\frac{ k^4 \tau_R^4+i k^2 \omega \tau_R^3 \left(R+i \tau_R \omega +1\right)}{k^4 \tau_R^4+k^2 \omega\tau_R^3  (-\omega\tau_R +2 i)-4 i \omega^3\tau_R^3}\right),\\
   G^{02,02}_{TT} &=-\frac{(\varepsilon_0+P_0) \left(k^2 \tau_R^2-2 \omega\tau_R  \left(-i R+\omega\tau_R +i\right)\right)}{k^2 \tau_R^2-4 i \omega\tau_R },\label{cor-2d-rom0202}\\
G^{22,22}_{TT}&=(\varepsilon_0+P_0)\notag\\
&\times\left(1-\frac{\left(k^2\tau_R^2-\omega ^2\tau_R^2\right) \left(k^4 \tau_R^4+2 \omega ^3\tau_R^3 \left(-i R+ \omega\tau_R +i\right)+k^2 \omega\tau_R^3  \left(-3 \omega\tau_R +2 i R\right)\right)}{k^6 \tau_R^6+k^4 \omega\tau_R^5  (2 i-\omega\tau_R)-4 i k^2
   \omega ^3\tau_R^5}\right)\label{cor-2d-rom2222}.
\end{align}

\subsection{Finite temperature correlators with momentum relaxation}
\textit{Longitudinal relaxation.} In $3+1$ dimensions, the presence of momentum relaxation in the longitudinal direction only affects the spin $0$ channel:
\begin{align}\label{cor-3d-mombrTT}
G^{00,00}_{TT} &= -3(\varepsilon_0+P_0)\left(1+\frac{k^2 \tau_R  (2 k \tau_R+L (\tau_R \omega +i))}{-6 i k \tau_R
   \omega  \Gammat_\parallel +3 L \omega \Gammat_\parallel (1-i \tau_R \omega )+2 k^3 \tau_R^2+i k^2
   L \tau_R}\right),\\
   G^{22,00}_{TT} &= -(\varepsilon_0+P_0)\notag\\
   &\times \bigg(1+\frac{i \left(4 k^3 \tau_R^2+k^2 L \tau_R (3 \tau_R \omega +2 i)-6 k \tau_R \omega  (\tau_R \omega +i)-3 L \omega  (\tau_R \omega +i)^2\right)}{2 \left(6 k \tau_R \omega  \Gammat_\parallel+3 L \omega  \Gammat_\parallel (\tau_R \omega +i)+2 i k^3 \tau_R^2-k^2 L \tau_R\right)}\bigg),
\end{align}
where we introduced the dimensionless $\Gammat \equiv \Gamma\tau_R - 1$. In the $2+1$ dimensional case the even channel is modified to
\begin{align}
    G^{00,00}_{TT} &=-2 (\varepsilon_0+P_0)\notag\\
    &\times\left(1+\frac{ \left(k^4 \tau_R^4+i k^2 \omega\tau_R^3  \left(-2 \Gamma_\parallel \tau_R+R+i \tau_R \omega +1\right)\right)}{-k^2 \omega\tau_R^3  (4 i \Gamma_\parallel \tau_R+\tau_R
   \omega -2 i)-4 \omega ^2 \tau_R^2 \Gammat_\parallel (\Gamma_\parallel\tau_R-i \omega\tau_R )+k^4 \tau_R^4}\right), \label{cor-2d-mombr0000}\\
   G^{22,00}&=-2 (\varepsilon_0+P_0)\notag\\
   &\times\frac{ \left(k^2\tau_R^2- \omega^2\tau_R^2 -i \Gamma_\parallel\omega\tau_R^2\right) \left(k^2 \tau_R^2+i \omega\tau_R  \left(-2
   \Gamma_\parallel \tau_R+R+i \tau_R \omega +1\right)\right)}{-k^2 \omega\tau_R^3
    (4 i \Gamma_\parallel \tau_R+\tau_R \omega -2 i)-4 \omega ^2\tau_R^2 \Gammat_\parallel (\Gamma_\parallel\tau_R-i \omega\tau_R )+k^4
   \tau_R^4}.
\end{align}

\noindent \textit{Transverse relaxation.} In the $3+1$ dimensional case, the presence of transverse momentum relaxation affects the spin $1$ channel:
\begin{align}
G^{02,02}_{TT} &=3(\varepsilon_0+P_0)\bigg(-\frac{1}{3}\notag\\
&+\frac{ i \tau_R \omega  \left(L \left(-k^2 \tau_R^2+(\tau_R \omega +i)^2\right)+2
   k \tau_R (\tau_R \omega +i)\right)}{3 L \Gammat_\perp \left(k^2 \tau_R^2-(\tau_R \omega
   +i)^2\right)+2 k \tau_R \left(-3 \Gamma_\perp \tau_R (\tau_R \omega +i)+2 i k^2 \tau_R^2+3 \tau_R
   \omega +3 i\right)}\bigg),
\end{align}
while, analogously, in the $2+1$ dimensional case, it modifies the odd sector:
\begin{align}
    G^{02,02}_{TT} &=-2(\varepsilon_0+P_0)  \left(1-\frac{ i \omega\tau_R  \left(2 \Gamma_\perp \tau_R-R-i \tau_R \omega -1\right)}{-4 \Gamma_\perp^2 \tau_R^2+\Gamma_\perp\tau_R (4+4 i \tau_R \omega )+k^2 \tau_R^2-4
   i \omega\tau_R }\right).
\end{align}

\subsection{Finite temperature and finite density correlators}

{Next, we turn on both an external gauge field and the metric. In this case, the $\langle TT \rangle$ correlators remain unaffected by $\mu_0 \neq0$. This somewhat unintuitive result of the `asymmetric' representation of modes that exist in the spectra of $\langle TT\rangle$, $\langle JJ \rangle$ and $\langle TJ \rangle$ correlators is a consequence of the equation of state that we consider. In fact, the same phenomenon is also exhibited by the correlators computed purely from the theory of hydrodynamics. We show this in Appendix~\ref{app:canonicalapproach}.}

We therefore present only $\langle JJ \rangle$ and $\langle TJ \rangle$ correlators. The correlators in the $3+1$ dimensional case are 
\begin{align}\label{cor-3d-coup00}
    G^{0,0}_{JJ}&=\frac{\chi\left(2 k \tau_R + (\omega\tau_R + i)L\right)}{4\left(L - 2ik\tau_R\right)}\notag\\
    &\times\bigg(i + \frac{3 k^2 \tau_R^2\left(2ik\tau_R -L\right)}{2k^3\tau_R^3 + i k^2 \tau_R^2 L + 6i\omega k \tau_R^2 + 3i\omega\tau_R L\left(\omega\tau-R + i\right)}\bigg), \\
   \label{cor-3d-coup22}
   G^{2,2}_{JJ} &=\frac{\chi  \omega  \left(L \left(-k^2 \tau_R^2+(\tau_R \omega +i)^2\right)+2 k \tau_R (\tau_R \omega
   +i)\right)}{16 k^3 \tau_R^2}\notag\\
   &\times \left(1+\frac{12 k^3 \tau_R^3}{3 i L \left(k^2 \tau_R^2-(\tau_R \omega +i)^2\right)+2 k \tau_R \left(2 k^2
   \tau_R^2-3 i \tau_R \omega +3\right)}\right), \\
    G^{00,0}_{TJ}&= -\frac{3 k^2 T_0 \tau_R \chi  (2 k \tau_R+L (\tau_R \omega +i))}{i k^2 L \tau_R+2 k^3 \tau_R^2+6 i k \tau_R \omega +3 i L \omega  (\tau_R \omega +i)},
    \label{3d-gtj-000}\\
    G^{22,0}_{TJ} &= -\frac{T_0 \chi  \left(k^2 L \tau_R (3 \tau_R \omega +2 i)+4 k^3 \tau_R^2-6 k \tau_R \omega 
   (\tau_R \omega +i)-3 L \omega  (\tau_R \omega +i)^2\right)}{2( i k^2 L \tau_R+2 k^3 \tau_R^2+6 i k
   \tau_R \omega +3 i L \omega  (\tau_R \omega +i))},\label{3d-tj-220}\\
   G^{02,2}_{TJ} &= \frac{3 T_0 \tau_R \chi  \omega  \left(L \left(-k^2 \tau_R^2+(\tau_R \omega +i)^2\right)+2 k \tau_R
   (\tau_R \omega +i)\right)}{3 i L \left(k^2 \tau_R^2-(\tau_R \omega +i)^2\right)+2 k \tau_R \left(2 k^2
   \tau_R^2-3 i \tau_R \omega +3\right)}.\label{3d-tj-022}
\end{align}

In the $2+1$ dimensional case, this is given by
\begin{align}\label{cor-2d-coup00}
    G^{0,0}_{JJ}&=\chi\frac{  (R+i\omega\tau_R  -1)}{R-1}\frac{1}{3 R \left(k^2\tau_R^2 (R-1) +2 i (R-1) \omega\tau_R -2 \omega^2\tau_R^2\right)}\notag\\
    &\times\left(-3 k^4 \tau_R^4+k^2 \tau_R^2 (3 R+\omega\tau_R   (3\omega\tau_R +4 i)-3) +2 \omega\tau_R  ( \omega\tau_R +i) (R+i  \omega\tau_R -1) \right), \\
   \label{cor-2d-coup22}
   G^{2,2}_{JJ} &=\chi\frac{  \omega\tau_R  \left(k^2 \tau_R^2 \left(-3 i R+3 \omega\tau_R
   -i\right)+4 \omega\tau_R  \left(-R-i  \omega\tau_R +1\right)\right)}{3 k^2
   \tau_R^2 \left(k^2 \tau_R^2-4 i \omega\tau_R \right)},\\
   \label{cor-2d-coup000}
    G^{00,0}_{TJ}&=-2(\varepsilon_0+P_0)\frac{k^2\tau_R^2  \left(k^2 \tau_R^2+i \omega\tau_R  \left(R+i  \omega\tau_R +1\right)\right)}{3 T_0 \left(k^4 \tau_R^4+k^2 \omega\tau_R^3  (2 i- \omega \tau_R)-4 i
   \omega ^3\tau_R^3\right)},\\
   \label{cor-2d-coup220}
   G^{22,0}_{TJ} &=-\frac{(\varepsilon_0+P_0) \left(k^4 \tau_R^4 +2 \omega^3\tau_R^3 \left(-i R+ \omega\tau_R +i\right)+k^2 \omega \tau_R^3 \left(-3 \omega\tau_R +2 i R\right)\right)}{3 T_0 \left(k^4 \tau_R^4+k^2 \omega\tau_R^3  ( 2 i- \omega\tau_R)-4 i \omega ^3\tau_R^3\right)} , \\
      \label{cor-2d-coup022}
   G^{02,2}_{TJ}&=2(\varepsilon_0+P_0) \frac{\omega\tau_R  \left( \omega\tau_R -i \left(1+R\right)\right)}{3 T_0 \left(k^2 \tau_R^2-4 i \omega\tau_R \right)} .
    \end{align}
{We bring to the attention the factorization of the denominator of $G_{JJ}^{0,0}$ in the uncoupled cases (both in $2+1$ and $3+1$ dimensions) which are now effectively a product of the denominator of the \textit{uncoupled} correlators $G^{0,0}_{JJ}$ and $G^{00,00}_{TT}$. This factorization carries over to the cases with momentum dissipation that we consider next and motivates how we present the analysis part of the paper.}

\subsection{Finite temperature and finite density correlators with longitudinal momentum relaxation}

Finally, turning on momentum dissipation, the correlators in $3+1$ dimensions in response to an external gauge field and metric are
\begin{align}
    G^{0,0}_{JJ}&=\frac{\chi  (2 k \tau_R+L (\omega\tau_R  +i))}{4 (L-2 i k \tau_R)}\notag\\
    &\times\left(i-\frac{3 k^2 \tau_R^2 (2 k \tau_R+i L)}{6 k\omega \tau_R^2  \Gammat_\parallel+3 L   \Gammat_\parallel \omega\tau_R (\omega\tau_R +i)+2 i k^3 \tau_R^3-k^2\tau_R^2 L }\right),\label{cor-3d-mombr-00}\\
    G^{00,0}_{TJ}&=-\frac{3(\varepsilon_0+P_0) k^2 \tau_R^2  (2 k \tau_R+L ( \omega\tau_R +i))}{4 T_0 \left(-6 i k
    \omega\tau_R^2  \Gammat_\parallel+3 L \omega\tau_R  \Gammat_\parallel (1-i \tau_R \omega )+2 k^3
   \tau_R^3+i k^2\tau_R^2 L\right)},\\
   G^{22,0}_{TJ} &=-\frac{(\varepsilon_0+P_0) }{8 T_0 \left(-6 i k \omega\tau_R^2 
   \Gammat_\parallel+3 L \omega\tau_R  \Gammat_\parallel (1-i \omega\tau_R )+2 k^3 \tau_R^3+i k^2\tau_R^2 L
   \right)}\notag\\
   &\times\bigg(4 k^3 \tau_R^3+k^2\tau_R^2 L  (3 \omega\tau_R +2 i)-6 k\omega \tau_R^2 (\omega\tau_R  +i)-3 L \omega\tau_R  (\omega\tau_R +i)^2\bigg).
\end{align}
The same correlators as above, but in $2+1$ dimensions are given by
\begin{align}\label{cor-2d-JJ-mombr}
    G^{0,0}_{JJ}&=\frac{\chi  (R+i \tau_R \omega -1)}{R-1}\frac{1}{3R \left(k^2\tau_R^2 (R-1) +2 \omega\tau_R  \Gammat_\parallel (-i R+\tau_R \omega +i)\right)}\notag\\
    &\times\Big(k^2 \tau_R^2 (\tau_R \omega  (2 i \Gamma_\parallel \tau_R+3 \tau_R \omega +4 i)+3 R-3)\notag\\
    &~~~~~ - 2 \omega\tau_R \Gammat_\parallel (\tau_R \omega +i) (R+i \tau_R \omega -1)-3 k^4 \tau_R^4\Big),\\
    G^{00,0}_{TJ} &= \frac{ -2(\varepsilon_0+P_0)k^2 \tau_R^2\left(k^2 \tau_R^2+i \omega\tau_R  (-2 \Gamma_\parallel \tau_R+R+i \tau_R \omega
   +1)\right)}{3 T_0 \left(-k^2 \omega\tau_R^3  (4 i \Gamma_\parallel \tau_R+\tau_R \omega -2 i)-4 \omega ^2\tau_R^2 \Gammat_\parallel (\Gamma_\parallel\tau_R-i \omega\tau_R)+k^4 \tau_R^4\right)},\\
   G^{22,0}_{TJ} &=-\frac{(\varepsilon_0+P_0) }{3 T_0 \left(-k^2 \omega\tau_R^3  (4 i \Gamma_\parallel
   \tau_R+\omega\tau_R -2 i)-4 \omega^2\tau_R^2 \Gammat_\parallel (\Gamma_\parallel\tau_R-i \omega\tau_R )+k^4 \tau_R^4\right)}\notag\\
   &\times\left(i k^2 \omega\tau_R^3  (-2 \Gamma_\parallel \tau_R+2 R+3 i \omega\tau_R ) +2 \omega^2\tau_R^2
   (\Gamma_\parallel\tau_R-i \omega\tau_R ) (R+i \omega\tau_R -1)+k^4 \tau_R^4\right) .
\end{align}
\subsection{Finite temperature and finite density correlators with transverse momentum relaxation}
We start with the $3+1$ dimensional results:
\begin{align}
    G^{2,2}_{JJ} &= \frac{\chi  \omega\tau_R  \left(L \left(-k^2 \tau_R^2+(\omega\tau_R +i)^2\right)+2 k \tau_R ( \omega\tau_R
   +i)\right)}{16 k^3 \tau_R^3}\notag\\
   &\times\bigg(1+\frac{12 i k^3 \tau_R^3}{3 L \Gammat_\perp \left(k^2 \tau_R^2-(\omega\tau_R +i)^2\right)+2 k
   \tau_R \left(-3 \Gamma_\perp \tau_R (\omega\tau_R +i)+2 i k^2 \tau_R^2+3 \omega\tau_R +3
   i\right)}\bigg),\\
  G^{02,2}_{TJ}&=\frac{3 i \tau_R \omega  (\varepsilon_0+P_0) \left(L \left(-k^2 \tau_R^2+(\omega\tau_R +i)^2\right)+2
   k \tau_R (\omega\tau_R +i)\right)}{4 T_0 \left(3 L \Gammat_\perp \left(k^2 \tau_R^2-(\omega\tau_R +i)^2\right)+2 k \tau_R \left(-3 \Gamma_\perp \tau_R (\omega\tau_R +i)+2 i k^2
   \tau_R^2+3 \omega\tau_R +3 i\right)\right)}.
\end{align}

Finally, in $2+1$ dimensions, we have
\begin{align}
    G^{2,2}_{JJ} &= \frac{\chi  \omega\tau_R  \left(R (\omega\tau_R +i)-i \left(k^2 \tau_R^2-( \omega\tau_R +i)^2\right)\right)}{3 k^2
   \tau_R^2 R}\notag\\
   &\times\bigg(1+\frac{2 k^2\tau_R^2 R}{R \left(2 i \Gamma_\perp \tau_R ( \omega\tau_R +i)+k^2 \tau_R^2-2 i  \omega\tau_R +2\right)+2 \Gammat_\perp \left(k^2 \tau_R^2-( \omega\tau_R +i)^2\right)}\bigg),\\
   G^{02,2}_{TJ}&=\frac{2 i (\varepsilon_0+P_0)  \omega\tau_R  \left(k^2 \tau_R^2+R (-1+i\omega\tau_R )-(\omega\tau_R +i)^2\right)}{3 T_0 \left(R \left(2 \Gamma_\perp \tau_R (1-i  \omega\tau_R )-k^2 \tau_R^2+2 i
   \omega\tau_R -2\right)-2 \Gammat_\perp \left(k^2 \tau_R^2-( \omega\tau_R +i)^2\right)\right)}.
\end{align}

\section{Finite temperature results}\label{sec:charge-neutral}

In this section, we consider the physics implied by the correlators at finite temperature ($T \neq 0$), zero charge density ($\mu = 0$) and with unbroken conservation of momentum ($\Gamma = 0$).

In order to systematically analyze the results, we categorize the perturbations (and therefore correlators) according to the transformation properties of the symmetry group of the perpendicular space. In the case of $3+1$ dimensions this involves considering the $SO(2)$ group acting on the transverse plane, dividing the perturbations into scalars, vectors, and tensors. In the $2+1$ dimensions we have the $\mathbb{Z}_2$ symmetry $y\to-y$, which allows us to classify our objects into odd and even sectors. For a summary of the structures of the spectra, see Tables \ref{table-3plus1} and \ref{table-2plus1}.

\subsection{$3+1$ dimensions}\label{subsec:3+1-uncharged}

The spectrum of the energy-momentum tensor and current correlators at $T \neq 0$ (and $\mu = \Gamma = 0$ was analyzed by Romatschke in 2015 \cite{Romatschke:2015gic}. As this is the simplest case, we begin with the recap of those results. 

First, we note that due to the presence of the logarithm, there is a pair of branch points in all of the correlators at
\begin{align}
    \omega(k) =-\frac{i}{\tau_R}\pm k.
\end{align}
The pole structures in each of the cases can be summarized as follows.
\begin{itemize}
\item $\langle JJ\rangle$:

\textbf{Spin 0.} Considering the analytic structure of \eqref{cor-3d-rom00}, we see that there is a single pole at
\begin{align}\label{eq:DiffusiveMode}
    \omega(k) = \frac{i}{\tau_R} \left(k \tau_R \cot (k \tau_R)-1\right). 
\end{align}
This exact dispersion relation can be written in terms of the hydrodynamic (gradient expanded) power series as (see Refs.~\cite{Grozdanov:2019kge,Grozdanov:2019uhi})
\begin{align}\label{Diff_ser_T}    
    \omega(k) = - i \sum_{n=1}^\infty a_n k^{2n}, \quad a_n = - \frac{(-4)^n B_{2n} \tau_R^{2n-1}}{(2n)! },
\end{align}
where $B_n$ are the Bernoulli numbers. 

\textbf{Spin 1.} The correlator \eqref{cor-3d-rom22} has no additional analytic structure apart from the standard branch cut present in all correlators.
\end{itemize}

\noindent
\textit{Transport coefficients.} We can directly use the hydrodynamic series to determine the charge diffusion constant by expanding 
\begin{equation}
    \omega(k) = -i D k^2 + \mathcal{O}(k^4).
\end{equation}
Hence, $D$ is given in terms of the relaxation time by 
\begin{equation}
    D=\frac{\tau_R}{3}.
\end{equation}
Another way to extract the transport coefficients (which we will use in most of the work) is to use the explicit correlators along with the Kubo formulae (see Ref.~\cite{Romatschke:2015gic}). For thermal charge transport, we have 
\begin{align}
    D&=-\frac{1}{\chi} \lim_{\omega\rightarrow0}\lim_{k\rightarrow0} \frac{\omega}{k^2}\text{Im}\,\,G_{JJ}^{0,0} = \frac{\tau_R}{3},\\
    \tau_\Delta &=\frac{1}{\chi D} \lim_{\omega\rightarrow0}\lim_{k\rightarrow0} 
    \frac{1}{k^2}
    \text{Re}\,\,G_{JJ}^{0,0}=\tau_R,
\end{align}
where $\tau_\Delta$ is a second-order transport coefficient.
\begin{itemize}
\item $\langle TT \rangle$:\\
\textbf{Spin 0.} Turning our attention to the spin 0 channel \eqref{cor-3d-rom0000}, we find a pair of sound modes
\begin{equation}\label{eq:SoundMode-3d-uncoup}
\omega(k) = \pm\frac{1}{\sqrt{3}}k -  \frac{2i}{15} \tau_R\, k^2  +\mathcal{O}(k^3).
\end{equation}
\textbf{Spin 1.} Next, the spin 1 channel is diffusive
\begin{align}\label{spin1}
    \omega(k) = -i \frac{\tau_R}{5}k^2+\mathcal{O}(k^4).
\end{align}
\textbf{Spin 2.} Finally, we have the spin 2 channel \eqref{spin-2-thermal}, which has no additional pole structure, just like the spin 1 channel of $\langle JJ \rangle$.
\end{itemize}
\noindent
\textit{Transport coefficients.} We can extract the following transport coefficients
\begin{align}
    \eta &= -\lim_{\omega\rightarrow0}\lim_{k\rightarrow0} \frac{1}{\omega}\text{Im}\,G_{TT}^{23,23}=\frac{\tau_R T_0 s}{5},\\ \tau_\Pi&=\frac{1}{\eta}\lim_{\omega\rightarrow0}\lim_{k\rightarrow0} \left(\frac{1}{\omega^2}\text{Re}\,G_{TT}^{23,23}+\frac{\kappa}{2}\right)=\tau_R,\\
\kappa&=-2\lim_{\omega\rightarrow0}\lim_{k\rightarrow0} \frac{1}{k^2}\text{Re}\,G_{TT}^{23,23}=0,
\end{align}
where $\eta$ is the shear viscosity, $s=(\varepsilon_0+P_0)/T_0$ is the entropy density, and $\tau_\Pi$ and $\kappa$ are hydrodynamic second-order transport coefficient. They are the only two coefficients that play a role in linearized conformal hydrodynamics and thereby enter into two-point functions (see Ref.~\cite{Baier:2007ix}). In this work, for brevity, we will not explicitly study any higher-order transport coefficients (see Refs.~\cite{Grozdanov:2015kqa,Jaiswal:2013vta,deBrito:2023tgb,Diles:2019uft,Diles:2023tau}), although, since we have access to explicit Green's functions for all $\omega$ and $k$, we could compute them at any order in the gradient expansion.

\subsection{$2+1$ dimensions}\label{subsec:2+1-noncharged}

We now turn our attention to the analytic structure of the correlators in $2+1$ dimensions. We note that there is also a branch cut, however, it arises from a square root, instead of a logarithm as in $3+1$ dimensions. The pair of branch points is located at the same values of momenta as in $3+1$ dimensions, i.e. at 
\begin{align}
    \omega(k) = -\frac{i}{\tau_R}\pm k.
\end{align}
 
\begin{figure}[ht!]
    \centering
    \includegraphics[width=\textwidth]{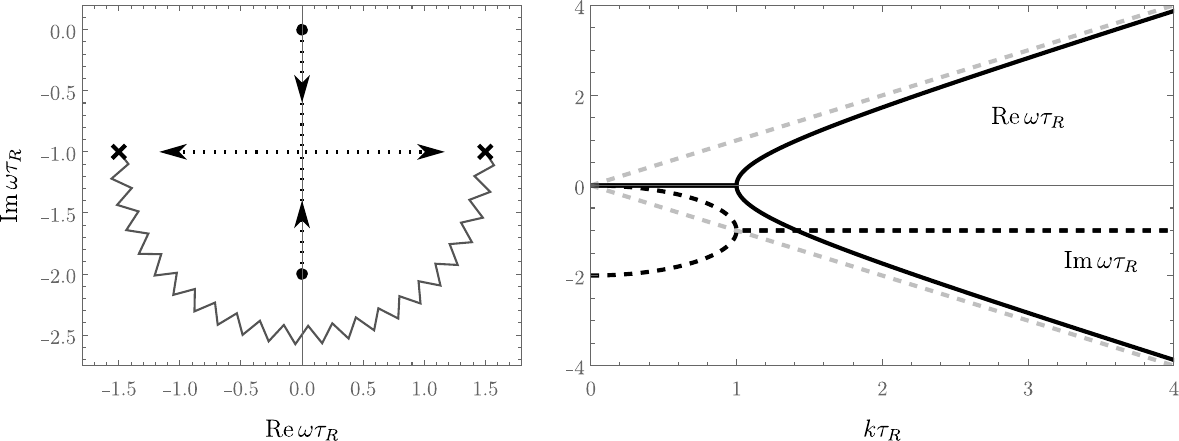}
    \caption{Left: The analytic structure of the correlator $G_{JJ}^{0,0}$ from \eqref{cor-2d-uncoup00} is plotted in the complex frequency plane. All other parameters set to unity. The positions of the two poles of \eqref{2plus1-poles} are indicated by dotted lines, while the arrows represent the movement direction as $0<k\tau_R<1.5$. We plot the branch cut at $k\tau_R=1.5$. The black crosses represent the branch points while the chosen branch cut is shown as a jagged line. Right: Real and imaginary part of the poles \eqref{2plus1-poles} in the $2+1$ dimensional case. The faint dashed grey lines represent the asymptotes $\pm k\tau_R$ of the real part of the poles.}\label{fig:poles2plus1JJ}
\end{figure}

\begin{itemize}
\item $\langle JJ \rangle$:

\textbf{Even.} For the correlator \eqref{cor-2d-uncoup00}, the poles are given by 
\begin{align}\label{2plus1-poles}
    \omega(k) = \frac{-i\pm\sqrt{k^2 \tau_R^2-1}}{\tau_R},
\end{align}
which we plot in the right panel of Figure~\ref{fig:poles2plus1JJ}. We therefore find a gapless pole exhibiting charge diffusion and gapped pole. Expanded as power series, their dispersion relations are, respectively,
\begin{align}
    \omega (k) &= - \frac{i}{2} \tau_R k^2 + \mathcal{O}(k^4), \\ 
    \omega (k) &= - \frac{2i}{\tau_R} + \frac{i}{2} \tau_R k^2 + \mathcal{O}(k^4). 
\end{align}
The poles exhibit a collision for real $k$, which is qualitatively of type described by the theory of quasihydrodynamics \cite{Grozdanov:2018fic}. The collision occurs at $k_*=1/\tau_R$ and for real $k>k_*$, the dispersion relations acquire a real part and thereby correspond to propagating modes. We present the collision in Figure~\ref{fig:poles2plus1JJ}. As is pointed out in Appendix~\ref{app:ComplexAnalysis}, the two poles \eqref{2plus1-poles} must be on the same sheet as they come from an equation with a chosen and fixed sheet of the square root (in this case the sheet with non-negative real part) given by the choice of integration contour in \eqref{eq:AngularIntegral}.

\textbf{Odd.} As was the case in $3+1$ dimensions, the odd channel does not exhibit additional structure apart from the branch cut.

\end{itemize}

\noindent
\textit{Transport coefficients.} The diffusion coefficient and the second order hydrodynamic timescale can be extracted  from $G^{0,0}_{JJ}$ \cite{Romatschke:2015gic} to find
\begin{align}
    D&= \frac{\tau_R}{2}, \\ \tau_\Delta &= \tau_R.
\end{align}
\begin{itemize}
\item $\langle TT \rangle$:

 \textbf{Even.} Next, we consider the correlator \eqref{cor-2d-rom0000}. This correlator has a sound mode
 \begin{equation}\label{eq:SoundMode-2d-uncoup}
\omega(k) = \pm\frac{1}{\sqrt{2}}k -  \frac{i}{8} \tau_R\, k^2  +\mathcal{O}(k^3).
\end{equation}
The third zero of the denominator in \eqref{cor-2d-rom0000} has a positive imaginary part. However, it does not represent a pole as the correlator is analytic at that point.

\textbf{Odd.} The odd channel \eqref{cor-2d-rom0202} exhibits a diffusive pole with the (exact) dispersion relation
\begin{equation}\label{2d-spin1-diffusion}
    \omega(k) = - i \frac{\tau_R}{4}k^2.
\end{equation}
We note that this is an interesting and unexpected result of the calculation, one not generically expected from hydrodynamics. 
\end{itemize}
\noindent
\textit{Transport coefficients.} To compute the shear viscosity, we turn on only time dependent perturbations in the transverse direction, namely $\delta g_{12}=\delta g_{12}(t)$  \cite{Baier:2007ix}. The shear viscosity in this case is given by 
\begin{align}
    \frac{\eta}{s} &=\frac{\tau_R T_0}{4}
.\end{align}
 
\section{Finite temperature results with momentum dissipation}\label{sec:uncharged-mombr}

In this section, we consider finite temperature correlators with zero charge density and unbroken momentum conservation. This will only affect the $\langle TT \rangle$ correlators, so we limit our discussion to these cases. Furthermore, adding longitudinal momentum dissipation will affect only the spin $0$ channel in $3+1$ and even channel in $2+1$ dimensions. Similarly, adding transverse dissipation modifies the spin $1$ and odd sector in $3+1$ and $2+1$ dimensions, respectively. As outlined in the previous section, the logarithmic and square root branch cut will be present in the $3+1$ and $2+1$ dimensional cases, respectively.

\subsection{$3+1$ dimensions}\label{subsec:3+1-mombr}

\begin{itemize}
    \item $\langle TT \rangle$ with $\Gamma_{\parallel}\neq0$:\\
\textbf{Spin 0.} Let us consider the correlator \eqref{cor-3d-mombrTT} in the case of $k\sim\Gamma_\parallel \ll 1/\tau_R$. This is reminiscent of the coherent regime in \cite{Davison:2014lua} (cf. \cite{hartnoll2018holographic}) if we identify their $\Lambda$ with our inverse relaxation time $1/\tau_R$. As a result, we obtain two modes

\begin{align}\label{eq:3dCollidingPolesDispersion}
    \omega(k)  = \frac{\pm\sqrt{3}\sqrt{4 k^2 - 3 \Gamma_\parallel^2 }- 3 i \Gamma_\parallel}{6} + \frac{k^2\tau_R}{15}\left(\frac{\mp 2 \sqrt{3}\Gamma_\parallel}{\sqrt{4k^2 -3 \Gamma_\parallel^2}} - 2 i \right)+\mathcal{O}\left(k^4, \Gamma_\parallel^4 \right).
\end{align}
The above expression reduces to the calculated sound modes of \eqref{eq:SoundMode-3d-uncoup} in the limit of $\Gamma_\parallel \rightarrow 0$. We observe the collision of poles at $k_* = \sqrt{3}\Gamma_\parallel/2$, which allows us to conclude that propagating `sound-like' modes dictate the transport properties for $k>\sqrt{3}\Gamma_\parallel/2$. For $k<\sqrt{3}\Gamma_\parallel/2$ only dissipative `diffusive' hydrodynamic and gapped modes are present.

    \textbf{Spin 1 and 2.} These channels remain unchanged from the $\Gamma_\parallel=0$ case.
    
    \item $\langle TT \rangle$ with $\Gamma_{\perp}\neq0$: \\
    \textbf{Spin 0 and 2.} The analytic structure of these channels are unaffected by $\Gamma_\perp.$\\
    \textbf{Spin 1.} In this channel there is gapped `diffusive' mode due to the presence of $\Gamma_\perp$:
    \begin{align}\label{eq:3d-shifted-spin1}
        \omega(k) = - i \Gamma_\perp-i \frac{\tau_R}{5(1-\Gamma_\perp \tau_R)} k^2 + \mathcal{O}(k^4).
    \end{align}
    Taking the limit $\Gamma_\perp\rightarrow0,$ we recover \eqref{spin1}.
\end{itemize}
\subsection{$2+1$ dimensions}\label{subsec:2+1-mombr}

\begin{figure}
    \centering
    \includegraphics[width=\textwidth]{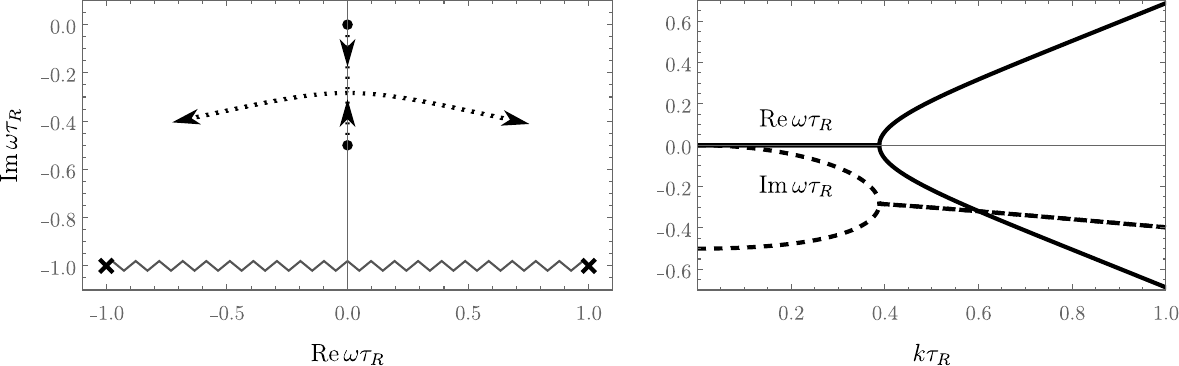}
    \caption{Left: the analytic structure of $T,\Gamma_\parallel\neq0$ $\langle TT \rangle$ correlator \eqref{cor-2d-mombr0000} in $2+1$ dimensions as a function of $10^{-4}<k \tau_R<1$ for $\Gamma_\parallel \tau_R=0.5$. For completeness, we include the branch cut evaluated at $k\tau_R=1$. There is a collision at $k\tau_R\approx 0.39$. Right: The dispersion relations of the same correlator.}
    \label{fig:2dTT-mombr}
\end{figure}

\begin{figure}
    \centering
    \includegraphics[width=\textwidth]{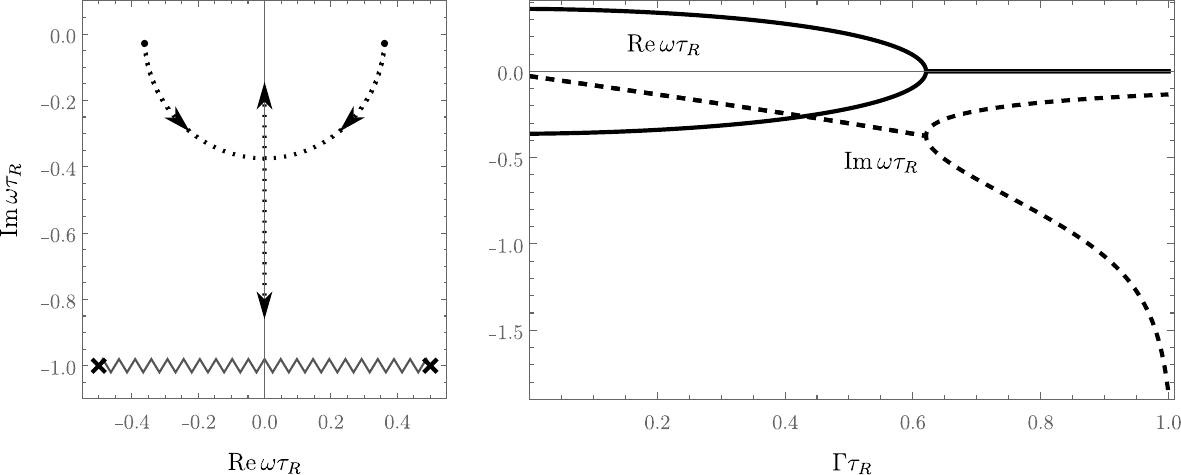}
    \caption{Left: same correlator as Fig.~\ref{fig:2dTT-mombr} as a function of  $10^{-4}<\Gamma\tau_R<0.8$ for $k\tau_R = 0.5$. Other parameters are set to unity. The poles collide at $\Gamma_* \tau_R \approx 0.62$.  Right: Plots of the dispersion relations of the poles.}
    \label{fig:2dTT-mombr-fixed-k}
\end{figure}

\begin{itemize}
    \item $\langle TT \rangle$ with $\Gamma_{\parallel}\neq0$: \\
    \textbf{Even.} In this channel, for small $k$, we have two purely damped `diffusive' modes, which collide for $k\sim \Gamma_\parallel$ and become propagating. We expand in $k\tau_R,\,\Gamma_\parallel \tau_R\ll 1$ and obtain
   \begin{align}\label{eq:2dCollidingPolesDispersion}
    \omega(k)  &= \frac{\pm\sqrt{2k^2 -\Gamma_\parallel^2} - i \Gamma_\parallel}{2} + \frac{\tau_R k^2}{8}\left(\mp\frac{\Gamma_\parallel\tau_R}{\sqrt{2k^2\tau_R^2 -\Gamma_\parallel^2\tau_R^2}} - i \right)
    +\mathcal{O}\left(k^4, \Gamma_\parallel^4 \right).
    \end{align}
The expressions \eqref{eq:2dCollidingPolesDispersion} reduce to the calculated sound modes of \eqref{eq:SoundMode-2d-uncoup} in the limit of $\Gamma_\parallel \rightarrow 0$. The poles now collide at $k_* = \sqrt{2}\Gamma_\parallel/2$. We show the collision of these poles as a function of $k \tau_R$ in Fig.~\ref{fig:2dTT-mombr} and as a function of $\Gamma \tau_R$ in Fig.~\ref{fig:2dTT-mombr-fixed-k}.\\
    \textbf{Odd.} This channel remains unchanged from the $\Gamma_\parallel=0$ case.
    \item $\langle TT \rangle$ with $\Gamma_{\perp}\neq0$:\\
    \textbf{Even.} The sound modes are unchanged and have the dispersion relations stated in \eqref{eq:SoundMode-2d-uncoup}.\\
    \textbf{Odd.} The `diffusive' mode is shifted by $\Gamma_\perp$
        \begin{align}\label{2d-diff-mode}
        \omega(k) = - i \Gamma_\perp-i \frac{\tau_R}{4(1-\Gamma_\perp \tau_R)} k^2+\mathcal{O}(k^4).
    \end{align}
\end{itemize}

\section{Finite temperature and finite density results}\label{sec:charged}

In this section, we turn our attention to the full thermoelectric case with $T \neq 0 $ and $\mu \neq 0$, still with momentum conservation ($\Gamma = 0 $). We first point out that the analytic structure of the $\langle TT\rangle$ correlators is not affected by switching on the external electric field, which is necessary for incorporating the effects of finite density. Hence, we will focus on the $\langle JJ \rangle$ and $\langle TJ\rangle$ correlators. The transport coefficients, with the exception of the thermoelectric ones, remain unchanged from the uncoupled case. Furthermore, the correlators in the full case inherit the branch cut structure discussed in Section~\ref{sec:charge-neutral}. Moreover, we note that for certain correlators, the analytic structure is factorizable into diffusive and sound modes, arising from $\langle JJ \rangle$ and  $\langle TT \rangle$, respectively.

Additionally, we can compute the thermoelectric conductivities \eqref{kubo}--\eqref{kubo_kappa}, which we find to be \cite{Hartnoll:2007ih,Hartnoll:2016apf}
\begin{align}
    \sigma&= \sigma_Q - \frac{1}{i\omega}\frac{n_0^2}{\varepsilon_0+P_0},\\
    \alpha=\tilde{\alpha}&=-\frac{\mu}{T_0}\sigma_Q- \frac{1}{i\omega T_0}\frac{n_0 s_0}{\varepsilon_0+P_0},\\
    \tilde{\kappa}&=\frac{\mu^2}{T_0}\sigma_Q- \frac{1}{i\omega T_0}\frac{T_0 s_0^2}{\varepsilon_0+P_0}, 
\end{align}
where $\sigma_Q$ is the DC finite part:
\begin{align}\label{sigmaQ}
    \sigma_Q(\omega=0) = \frac{\tau_R\,\chi}{d\,(d+1)}=\tau_R \frac{e^{\mu_0/T_0}}{12} \frac{T_0^{d-1}}{\pi^{d-1}}.
\end{align}
Here, $n_0$ is the equilibrium number density and $s_0 = (\varepsilon_0+P_0-\mu_0 n_0)/T_0$ is the equilibrium entropy density.
The Ward identities imply that the conductivities satisfy
\begin{align}
    (\mu_0 \sigma+T_0 \alpha)i \omega&=-n_0,\\
    (\tilde{\kappa}+\mu_0 \alpha)i \omega&=-s_0.
\end{align}

\subsection{$3+1$ dimensions}

\begin{figure}
\centering
    \includegraphics[width=\linewidth]{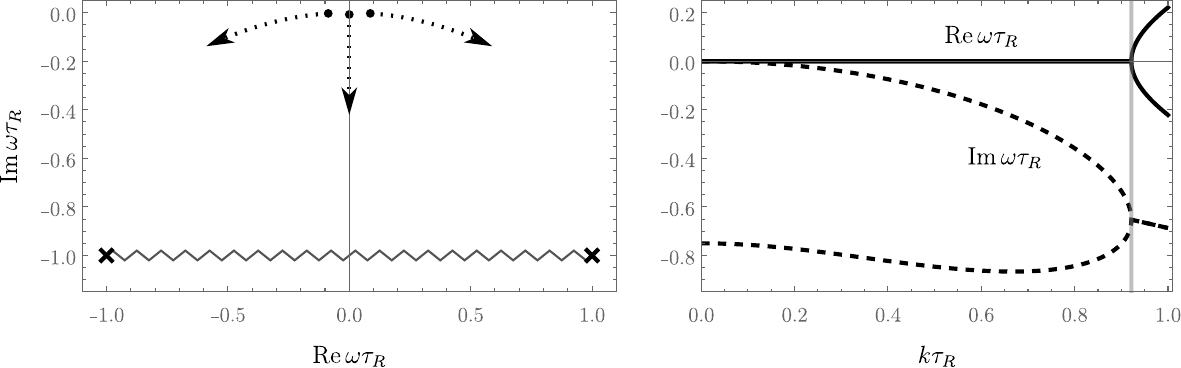}
    \caption{
   Left: The analytic structure of the $G^{0,0}_{JJ}$ correlator \eqref{cor-3d-coup00} for $0.15<k\tau_R<1$, with other parameters set to unity. The position of the poles \eqref{eq:DiffusiveMode} and \eqref{eq:SoundMode-3d-uncoup} are denoted with black dotted lines. The branch cut is evaluated at the fixed value of $k\tau_R=1$. Right: Real and imaginary part of the two zeros of the correlator, which collide in a manner similar to poles in quasihydrodynamic collisions. The faint grey line is located at the collision of the zeros, $k_*\tau_R \approx 0.9215$.
    }
    \label{fig:00Structure}
\end{figure}

\begin{itemize}
\item $\langle JJ \rangle$:

\textbf{Spin $0$.} For the $\langle JJ \rangle$ correlator \eqref{cor-3d-coup00}, we see that there are now three poles, which correspond to the two sound poles \eqref{eq:SoundMode-3d-uncoup} and the diffusive pole is given by \eqref{eq:DiffusiveMode}. The transport coefficients computed in the previous section are unchanged.

The analytic structure is presented in Figure \ref{fig:00Structure}. We note that the zeros of the correlator collide with one another at $k_* \tau_R \sim 0.9215$, developing a non-zero real part post collision. This is akin to the collision of poles in quasihydrodynamics \cite{Grozdanov:2018fic,Soloviev:2022mte}.

\textbf{Spin $1$.} Additionally, there is the spin $1$ $\langle JJ \rangle$ correlator \eqref{cor-3d-coup22} with a single diffusive pole with the same behavior as \eqref{spin1}.

\item $\langle TJ \rangle$:

\textbf{Spin $0$.} The correlators \eqref{3d-gtj-000} and \eqref{3d-tj-220} have two sound poles as in the decoupled case \eqref{eq:SoundMode-3d-uncoup}. 

\textbf{Spin $1$.} The correlator \eqref{3d-tj-022} has the spin $1$ diffusive mode \eqref{spin1}.

\end{itemize}

\subsection{$2+1$ dimensions}\label{sec:2d-coup}
We move onto the study of correlators and their analytic structure in $2+1$ dimensions. 
\begin{itemize}
\item $\langle JJ \rangle$:

\textbf{Even.} We expect no different factorization than the one we found in the $3+1$ dimensional case. Indeed, the even channel exhibits the same quasihydrodynamic poles \eqref{2plus1-poles}, together with the $2+1$ dimensional odd sound modes of \eqref{eq:SoundMode-2d-uncoup}.

\textbf{Odd.} Analogously, the odd channel is composed of the odd diffusive mode of \eqref{2d-spin1-diffusion}.

\item $\langle TJ \rangle$:

\textbf{Even.} We observe the even sound modes of \eqref{eq:SoundMode-2d-uncoup}.

\textbf{Odd.} This channel exhibits the diffusive pole of \eqref{2d-spin1-diffusion}.

\end{itemize}

\section{Finite temperature and finite density results with momentum dissipation}\label{sec:charged-mombr}

\subsection{$3+1$ dimensions}
Here we briefly reiterate the analytic structure from the above sections. Due to the presence of the factorization of the analytic structure, we can already anticipate the analytic structure and the dispersion relations for all the sectors in this fully coupled case. For the sake of completeness we refer the reader to the relevant results in the previous sections.
We consider only the $\langle JJ \rangle$ and $\langle TJ\rangle$ as the $\langle TT \rangle$ correlators remain unaffected by the coupling.

Furthermore, we point our that the conductivities in the presence of momentum breaking have a slightly different form \cite{Hartnoll:2007ih,Hartnoll:2016apf}:
\begin{align}\label{cond-br}
    \sigma=\sigma_Q-\frac{1}{i\omega-\Gamma}\frac{n_0^2}{\varepsilon_0+P_0},
\end{align}
where $\sigma_Q$ was defined in \eqref{sigmaQ}.
The Ward identities in the case of momentum breaking satisfy:
\begin{align}\label{eq:WardConductivities}
    (\mu_0 \sigma+T_0 \alpha)(i \omega-\Gamma)=-n_0,\quad (\bar{\kappa}+\mu_0 \alpha)(i \omega-\Gamma)=-s_0,
\end{align}
and the Onsager relations hold, i.e.,  $\alpha=\tilde{\alpha}$.

\begin{itemize}
    \item $\langle JJ\rangle$ with $\Gamma_\parallel\neq0$:\\
    \textbf{Spin 0.} The analytic structure of \eqref{cor-3d-mombr-00} is composed of the spin $0$ sector of uncharged $\langle JJ \rangle$ correlator discussed in Section~\ref{subsec:3+1-uncharged} and the spin $0$ sector with of $\langle TT \rangle_{\Gamma_\parallel \neq 0}$ discussed in Section~\ref{subsec:3+1-mombr}. The diffusive mode is therefore given by \eqref{eq:DiffusiveMode}, while the colliding poles are described by \eqref{eq:3dCollidingPolesDispersion}.  \\
    \textbf{Spin 1.} This sector is unaffected by the momentum dissipation. Since spin $1$ of $\langle JJ \rangle$ does not exhibit any poles, the only pole structure is therefore given by the spin $1$ channel of $\langle TT \rangle$, discussed in Section~\ref{subsec:3+1-uncharged}. The dispersion relation of the spin $1$ diffusive mode is given by \eqref{spin1}.\\
    \item $\langle TJ\rangle$ with $\Gamma_\parallel\neq0$:\\
    \textbf{Spin 0.} Here the channel inherits the analytic structure of the spin $0$ channel of $\langle TT \rangle_{\Gamma_\parallel \neq 0}$ discussed in Section~\ref{subsec:3+1-mombr}. The pole structure here is composed of the colliding poles with dispersion relation given by \eqref{eq:3dCollidingPolesDispersion}.\\
    \textbf{Spin 1.} This sector is unaffected by the longitudinal momentum dissipation. The only pole structure is therefore given by the spin $1$ channel of $\langle TT \rangle$, discussed in Section~\ref{subsec:3+1-uncharged}. The dispersion relation of the spin $1$ diffusive mode is given by \eqref{spin1}.\\

    \item $\langle JJ \rangle$ with $\Gamma_\perp\neq0$:\\
    \textbf{Spin 0.}  The analytic structure here is again a product of the uncoupled $\langle JJ \rangle$ spin $0$ channel and the spin $0$ channel of $\langle TT \rangle$ discussed in Section~\ref{subsec:3+1-uncharged}, since $\Gamma_\perp \neq 0$ affects only the spin $1$ channel. Consequently, we have two diffusive modes; the charge diffusion \eqref{eq:DiffusiveMode} and the spin $1$ diffusion \eqref{spin1} \\
    \textbf{Spin 1.} The pole structure of this sector is inherited by the spin $1$ $\langle TT \rangle_{\Gamma_\perp \neq 0}$ discussed in Section~\ref{subsec:3+1-mombr}. As such it exhibits the shifted diffusive-like mode of \eqref{eq:3d-shifted-spin1}.\\
    \item $\langle TJ \rangle$ with $\Gamma_\perp\neq0$:\\
    \textbf{Spin 0.} Setting $\Gamma_\perp \neq 0$ does not affect the spin $0$ channel. Therefore the analytic structure the same as the analytic structure of spin $0$ $\langle TT \rangle$ discussed in Section~\ref{subsec:3+1-uncharged} and the poles correspond to the two sound modes of \eqref{eq:SoundMode-3d-uncoup}.\\
    \textbf{Spin 1.} The analytic structure here is inherited from the spin $1$ sector of $\langle TT\rangle_{\Gamma_\perp\neq 0}$ discussed in Section~\ref{subsec:3+1-mombr}. The pole structure is composed of the shifted diffusive-like mode of \eqref{eq:3d-shifted-spin1}.

\end{itemize}

\noindent
    \textit{Transport coefficients}. Analogously to the longitudinal case, now the transverse coefficient of the diagonal conductivity matrix $\sigma$ obtains a Drude peak. The conductivity $\sigma^{22}$ is of the same form as \eqref{cond-br} with $\Gamma\rightarrow\Gamma_\perp$.

\subsection{$2+1$ dimensions}
\begin{figure}[hbp]
    \centering
    \includegraphics[width=\textwidth]{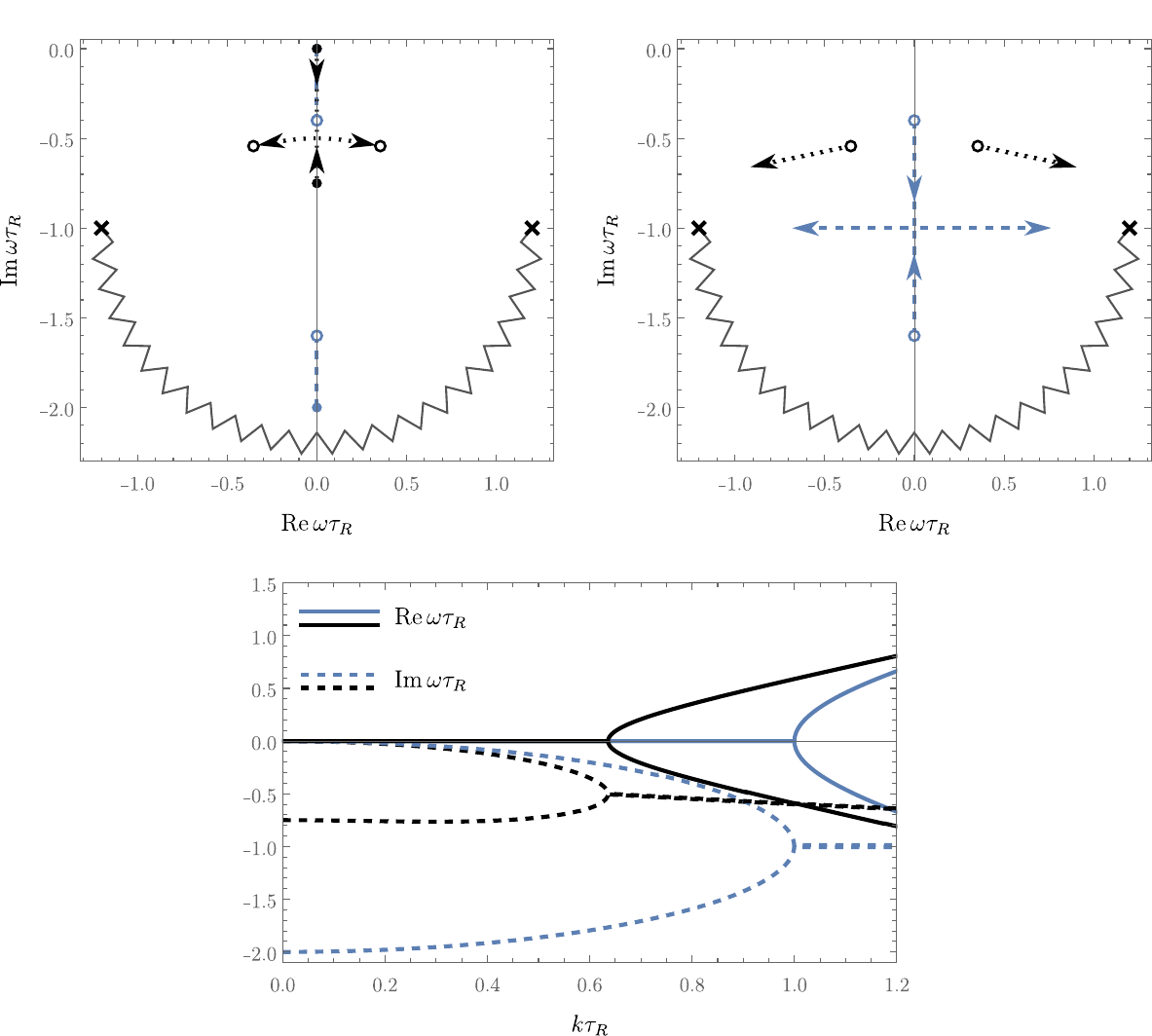}
    \caption{Top left: the analytic structure of the even $\langle JJ\rangle _{T, \mu,\Gamma_\parallel\neq0}$ correlator, given explicitly in \eqref{cor-2d-JJ-mombr}, as a function of $10^{-4}<k\tau_R<0.8$ for $\Gamma\tau_R = 3/4$. The blue lines denote the first pair of colliding poles, while the black lines describe another pair of modes that exhibits a crossover from damped (`diffusive-like') to propagating (`sound-like') behavior (also found in the $\langle TT\rangle$ correlator, see e.g.~Fig.~\ref{fig:2dTT-mombr}). Dots represent the start of the $k$-dependent evolution and empty circles depict the end of the evolution at $k\tau_R=0.8$, with each pair plotted in their respective color. The branch cut is evaluated at fixed $k\tau_R=1$. Top right: The evolution of the same correlator from $0.8<k\tau_R<1.2$, where we observe the second collision. Note again that in both the left and right plots, the empty circles denote $k\tau_R=0.8.$ Bottom: Real and imaginary parts of the four poles' dispersion relations. One collision of poles occurs at $k\tau_R\sim 0.64$ and the second at $k\tau_R = 1$.}
    \label{fig:jj-TmuG}
\end{figure}

We now turn our attention to the correlators in $2+1$ dimensions. As in the previous subsection, much of the analytic structure is factorizable.
\begin{itemize}
    \item $\langle JJ\rangle$ with  $\Gamma_\parallel\neq0$:\\
    \textbf{Even.} The spectrum has the structure that combines those of $\langle JJ \rangle$ discussed in Section~\ref{subsec:2+1-noncharged} and $\langle TT\rangle_{\Gamma_\parallel \neq 0}$ discussed in Section~\ref{subsec:2+1-mombr}. The dispersion relations are given by the quasihydrodynamic modes of \eqref{2plus1-poles}, which collide at $k\sim 1/\tau_R$ and the colliding poles of \eqref{eq:2dCollidingPolesDispersion}, which collide at $k\sim \Gamma_\parallel$. As can be seen in Figure~\ref{fig:jj-TmuG}, due to the factorization we observe two independent collision of poles. \\
    \textbf{Odd.}  The structure here is unaffected by $\Gamma_\parallel \neq 0$ and is therefore given by the odd channel of $\langle TT\rangle$ discussed in Section~\ref{subsec:2+1-noncharged}. The only present mode is the diffusive mode given by \eqref{2d-spin1-diffusion}\\
    \item $\langle TJ\rangle$ with  $\Gamma_\parallel\neq0$:\\
    \textbf{Even.} This channel has the same quasihydrodynamic collision of poles as in \eqref{eq:2dCollidingPolesDispersion}, see Section~\ref{subsec:2+1-mombr}.\\
    \textbf{Odd. } This channel is unaffected by $\Gamma_\parallel$. See Section~\ref{sec:2d-coup}.

    \item  $\langle JJ\rangle$ with  $\Gamma_\perp\neq0$:\\
    \textbf{Even.} This channel is unaffected by $\Gamma_\perp$. We refer the reader to Section~\ref{sec:2d-coup} for the analysis of \eqref{cor-2d-coup00}. \\
    \textbf{Odd.} This channel exhibits a diffusive-like mode, given by \eqref{2d-diff-mode}.\\
    \item $\langle TJ\rangle$ with  $\Gamma_\perp\neq0$:\\
    \textbf{Even.} This channel is unaffected by $\Gamma_\perp$. See Section~\ref{sec:2d-coup}. It retains its sound modes \eqref{eq:SoundMode-2d-uncoup}.\\
    \textbf{$\langle TJ\rangle$ Odd.} This channel develops a gap, with its diffusive-like mode given by \eqref{2d-diff-mode}.
\end{itemize}
\noindent
    \textit{Transport coefficients}. The conductivity matrix is now modified in the transverse component with $\sigma^{22}$ being of the form \eqref{cond-br} with $\Gamma \rightarrow\Gamma_\perp$.

\section{Conclusion}\label{sec:Conclusion}

In this work, we studied the linear response of a system of massless particles to external electric and metric perturbations in the RTA approximation of the kinetic theory Boltzmann equation. Owing to the theoretical elegance of the RTA, we were able to provide a complete classification of the analytic structure of the correlators of the theory, extending the computation of \cite{Romatschke:2015gic} to include momentum dissipation and the current-energy-momentum coupling at finite temperature and chemical potential, both in $3+1$ and $2+1$ dimensions. This is summarized in Tables \ref{table-3plus1} and \ref{table-2plus1}. We would like to emphasize that although we were considering linear response, the RTA includes non-linearities due to the non-trivial matching conditions. Furthermore, we extended the usual framework of RTA to explicitly include momentum dissipation.
The analysis we provided shows that the RTA correlators have a rich analytic structure with branch points, hydrodynamic modes and, in some cases, gapped modes. A number of spectra exhibit quasihydrodynamic collisions among their modes. Moreover, in the $\langle JJ \rangle$ correlators in $2+1$ dimensions, we also observed a pole collision that we claim is qualitatively the kinetic theory analogue of the `hydrodynamic-to-relativistic' crossover discussed in holography in  Ref.~\cite{Witczak-Krempa:2012qgh} (see also Refs.~\cite{Berti:2009kk,Grozdanov:2024wgo}).

There is a number of directions to explore from here. It would be interesting to implement the numerical methods of \cite{Ochsenfeld:2023wxz} to numerically explore the nonlinear structure of thermoelectric RTA. It would be interesting to see which features of the correlators persist moving away from the RTA limit by considering in greater detail, e.g. a $2\to 2$ scattering collision kernel.

The thermoelectric effect has physical applications in high energy systems, such as the quark gluon plasma (QGP) \cite{Abhishek:2020wjm,Das:2020beh,Kurian:2021zyb} and in condensed matter systems, such as cuprates and graphene \cite{Hartnoll:2007ih}. In stochastic hydrodynamic descriptions of the QGP, quasihydrodynamic modes are seen in models of the chiral phase transition \cite{Grossi:2020ezz,Grossi:2021gqi,Florio:2021jlx,Soloviev:2022mte,Florio:2023kmy}, indicating the presence of a phase transition from diffusive quark-like modes to propagating pion modes.
Furthermore, the QGP has been recently observed generating a large magnetic field \cite{STAR:2023jdd}, clearly indicating that such fields are important to understand. 
One limitation of the current work is the absence of magnetic fields, which are necessary to generate, e.g. the Hall conductivity \cite{Hartnoll:2007ai}.  As such, it would be instructive to see how the correlators change with the inclusion of fermionic degrees of freedom and other species, phase transitions and external magnetic fields. How this discussion would inform the kinetic correlators of magnetohydrodynamics remains to be seen.

It would also be worthwhile to understand how such correlation functions behave in expanding systems, especially in the context of  hydrodynamic attractors \cite{Soloviev:2021lhs}. Such attractors are characterized by the decay of non-hydrodynamic modes, leading to the system's approach to universal hydrodynamic behavior. The hope is that this would ultimately shed light on the underlying analytic structure of the quark gluon plasma \cite{Kurkela:2019kip}.

Finally, our intrinsically weakly coupled results (due to assumptions used in the construction of kinetic theory) also serve as a benchmark for comparison with strongly coupled holographic computations and allow for a further discussions of how qualitative and quantitative properties of spectra in quantum field theories transition from weak to strong coupling. Thereby, we hope this may serve as further motivation and encouragement for new detailed explorations of the analytic structures present in spectra of thermal field theories, for example in the style of Refs.~\cite{Waeber:2015oka,Grozdanov:2016vgg,Grozdanov:2016fkt,Solana:2018pbk,Grozdanov:2018gfx,Witczak-Krempa:2012qgh,Witczak-Krempa:2013xlz,Romatschke:2015gic,Kurkela:2017xis,Romatschke:2019gck,Romatschke:2019ybu,Romatschke:2019qbx}.

\begin{acknowledgments}
We would like to thank Danny Brattan, Nicolas Chagnet, Richard Davison, Blaise Gout{\' e}raux, Sean Hartnoll, Giuseppe Policastro, Paul Romatschke, Koenraad Schaalm and Mile Vrbica for fruitful discussions. The work of M.B. was supported by the project N1-0245 of Slovenian Research Agency (ARIS) and the financial support from the Slovenian Research Agency (ARIS) research core funding P1-0035. The work of S.G. was supported by the STFC Ernest Rutherford Fellowship ST/T00388X/1. The work is also supported by the research programme P1-0402 and the project N1-0245 of Slovenian Research Agency (ARIS). A.S. was supported by funding from Horizon Europe research and innovation programme under the Marie Skłodowska-Curie grant agreement No.~101103006 and the project N1-0245 of Slovenian Research Agency (ARIS).
\end{acknowledgments}

\appendix

\section{Branch cuts and Riemann sheets}\label{app:ComplexAnalysis}

In this Appendix, we discuss certain results from complex analysis that are relevant for the discussion of branch cuts, Riemann sheets and poles encountered in this work. In particular, we discuss how the ubiquitous branch cut with branch points $\omega = \pm k - i /\tau_R$ that arises from an angular integral can be deformed and what this means for the poles of the correlators. We show that the cut can be deformed arbitrarily within the same homotopy class and that all considered poles, which arise as solutions of equations that include multivalued functions (either of the complex logarithm or the square root), lie on the same sheet.

The branch cuts in our analysis arise from the angular integrals, which are generally proportional to
\begin{align}\label{eq:AngularIntegral}
    \int \frac{d\Omega}{\Omega_d} v^\mu v^\nu \dots \frac{\frac{f_0}{T_0} \tau_R \textbf{E}\cdot\textbf{v} - \frac{f_0}{T_0}\tau_R \Gamma^0_{\alpha \beta} p^0 v^\alpha v^\beta  
    + \delta f_{\text{eq}}}{1 + \tau_R (-i\omega + i \textbf{k}\cdot\textbf{v})},
\end{align}
where $v^\mu \equiv (1,\textbf{v})$. We will consider only the $3+1$ dimensional case without additional weights $v^\mu v^\nu \dots$ and note that the discussion and results in $2+1$ dimensions are completely analogous. The numerator of \eqref{eq:AngularIntegral} can be, through sequential use of integration by parts, set proportional to unity. Aligning the $z$-axis along the wavevector, we retrieve the following integral which is central to our results and discussion:
\begin{align}\label{eq:f_aIntegral}
    \int_{-1}^1 dx \frac{1}{a - x}\equiv \int_{-1}^1 d x f_a(x),
\end{align}
where $x\equiv \cos \theta$ and $a\equiv (i + \omega \tau_R)/(k\tau_R)$.\footnote{The integral in the $2+1$ dimensional case with $\textbf{k}=(k,0)$ would be $\int_0^{2\pi} d\phi\frac{1}{a - \cos\phi}$, which is qualitatively the same, but considerably less trivial to compute.} Keeping $\omega$ and $k$ real, we conclude that $\im a>0$. The integrand $f_a(x)$ is therefore analytic in $x$ in the lower half plane (LHP). Hence, any integration contour from $-1$ to $1$ that lies in the LHP gives rise to the same result as the integration over the `physical' contour, which runs over the real interval $[-1,1]$. This is because such paths are in the same homotopy class.

One might naturally identify the integral $\eqref{eq:f_aIntegral}$ with $\ln \frac{a-1}{a+1}$. However, keeping this logarithm completely general introduces an ambiguity in the choice of a branch, which would associate a multivalued map with a well-defined and single-valued integral. This can be remedied by considering all classes of (equivalent) integration contours, with each class containing all contours that can be smoothly (homotopically) deformed into each other. Indeed, it is precisely the different non-equivalent choices of the contour that encapsulate this multivaluedness, i.e., there is a one-to-one correspondence between non-equivalent integration contours and the sheets of the logarithm, given by the winding number of the contour with respect to $a$. 

While we are free to choose any path in the homotopy class of the physical contour, the resulting branch cut is \textit{not} independent of this choice. Formally, the integral does not exist when $a$ lies on the contour of integration (although one can compute the principle part of it), as the integration contour may not pass through a singularity. After the integration, we analytically continue $\omega\in\mathbb{C}$, which allows $a$ to take values in the whole complex plane. Since $a$ cannot move accross the integration contour, there is a natural `cut' in the complex $\omega$-plane given implicitly by
\begin{equation}
    a(\omega) \in \{\gamma(t);\,t\in[0,1]\},
\end{equation}
where $\gamma:[0,1]\rightarrow\mathbb{C}$ is our chosen integration contour. We conclude that the logarithm stemming from the integral \eqref{eq:f_aIntegral} is single-valued on the cut complex plane.  

We now turn our attention to the equations involving a square root in the $2+1$ dimensional cases. For example, the poles of the density-density correlator \eqref{cor-2d-uncoup00} are given by solutions to the equation
\begin{equation}\label{eq:2dPolesEq}
    \sqrt{k^2 \tau_R^2 - (\omega\tau_R+i)^2} = 1.
\end{equation}
Since the square root is a single-valued function on the cut complex plane, then, by definition, the argument must live on the cut complex plane and the square root must map to the same sheet. Therefore, the two poles are both present on the same sheet where the collision of poles discussed in the main body of the paper can occur. We note that the fact that the square root must map the two poles to the same sheet is clear from the equation \eqref{eq:2dPolesEq} itself, as the two sheets of the square root are composed by the two choices of the real part being either non-negative or non-positive. Solving equation \eqref{eq:2dPolesEq} explicitly demands that the real part of the square root is non-negative.

\section{Hydrodynamic correlators}\label{app:canonicalapproach}

{In this appendix, we compute all correlators analyzed in this work within the hydrodynamic approximation at $T \neq 0$, $\mu\neq0$ and with momentum dissipation. We do this using the so-called {\it canonical approach} (see Refs.~\cite{kadanoff,Kovtun:2012rj}).}

We start by perturbing the hydrodynamic fields in constitutive relations for the energy-momentum tensor and the current expanded to first order in the gradient expansion. In particular, we perturb the energy density $\varepsilon$, velocity field $u^i$ and charge density $n$ around their equilibrium values $\varepsilon_0$, $u^{i}_0 = 0$ and $n_0$ respectively.  To the first order in perturbations, in the Landau frame and in $d$ spatial dimensions \cite{Kovtun:2012rj}:
\begin{align}
    \delta T^\mn &= \delta\varepsilon u^\mu u^\nu +  (\varepsilon_0 + P_0) (\delta u^\mu u^\nu + u^\mu \delta u^\nu) + \delta P \Delta^\mn \notag \\
    &- \eta \Delta^{\mu\alpha}\Delta^{\nu\beta}\left(\nabla_\alpha \delta u_\beta + \nabla_\beta \delta u_\alpha - \frac{2}{d} g_\ab \nabla_\sigma \delta u^\sigma\right) - \zeta \Delta^\mn \nabla_\sigma \delta u^\sigma,\\
    \delta J^\mu &= \delta n u^\mu + n_0 \delta u^\mu - \sigma T_0 \Delta^{\mu\nu}\nabla_\nu \left(\frac{\delta \mu}{T_0} - \frac{\mu_0}{T_0}\frac{\delta T}{T_0}\right),
\end{align}
where $\eta$, $\zeta$ and $\sigma$ are shear viscosity, bulk viscosity and conductivity, respectively, and
\begin{equation}
    \Delta^\mn = g^\mn + u^\mu u^\nu.
\end{equation}

Next we write down the equations of motion for the hydrodynamic variables $\delta\varepsilon$, $\delta\pi^i$ and $\delta n$ arising from the conservation equations \eqref{eq:MomentumBreak} and \eqref{eq:ChargeConservation}. We restrict our discussion to flat spacetime and Fourier transform the spatial components choosing the wavevector ${\bf k}$ to align with the $x$ axis direction as in the main text: 
\begin{align}\label{canonical-system}
&\partial_t \delta \varepsilon + i k \delta \pi^x = 0,\\
&\partial_t \delta \pi^{\hat{i}} + ik \delta T^{x\hat{i}} = - \Gamma_{\hat{i}} \delta\pi^{\hat{i}},\\
&\partial_t \delta n + ik \delta J^x = 0,
\end{align}
where the hatted indices are not summed over. Note that the perpendicular channel decouples and satisfies a (damped) diffusion equation
\begin{equation}
    \partial_t \delta \pi^\perp + \frac{\eta}{\varepsilon_0 + P_0} k^2 \delta \pi^\perp = - \Gamma_\perp \delta \pi^\perp.
\end{equation}
Due to this decoupling we focus only on the coupled equations governing the evolution of $\delta \varepsilon$, $\delta \pi^\parallel$ and $\delta n$ and return to the perpendicular sector at a later time. 

Introducing $\varphi_a=(\delta\varepsilon, \delta \pi^{\parallel},\delta n)$, we may write \eqref{canonical-system} in a compact form:
\begin{align}\label{eq:M-systemEq}
    \partial_t \varphi_a+M_{ab}(k)\varphi_b=0.
\end{align}
The system \eqref{eq:M-systemEq} describes an initial value problem for which we provide the initial condition in the form $\varphi_a(t=0,\textbf{k}\rightarrow 0) = \chi_{ab}\lambda_b(\textbf{k}\rightarrow 0)$, where $\chi_{ab}$ is the static susceptibility matrix and $\lambda_b$ are the sources. For our particular case, the susceptibility matrix, relating the hydrodynamic variables $\varphi_a$ to their sources $\lambda_a=(\frac{\delta T}{T_0},\, \delta u^x,\, \delta \mu-\frac{\mu_0}{T_0} \delta T)$, is given by
\begin{align}\label{eq:hydroChi}
    \chi_{ab}&=\frac{\partial \varphi_a}{\partial \lambda_b}=\begin{pmatrix}
        T_0\frac{\partial \varepsilon}{\partial T} + \mu_0 \frac{\partial \varepsilon}{\partial \mu} & 0 & \frac{\partial \varepsilon}{\partial \mu}\\
        0 & \varepsilon_0 + P_0 & 0 \\
        T_0\frac{\partial n}{\partial T} + \mu_0 \frac{\partial n}{\partial \mu} & 0 & \frac{\partial n}{\partial \mu}
    \end{pmatrix}.
\end{align}

Solving the initial value problem \eqref{eq:M-systemEq} and transforming the solution via Laplace transform,\footnote{Our convention for the Laplace transform is $A(z,\textbf{k})=\int_0^\infty dt e^{izt}A(t,\textbf{k})$.} we obtain
\begin{align}\label{can-sol}
    \varphi_a(z,\textbf{k})= (K^{-1})_{ab}\chi_{bc}\lambda_c^0,
\end{align}
where
\begin{align}\label{eq:Kmatrix}
    K_{ab}&\equiv-i z\delta_{ab}+M_{ab}.
\end{align}
The matrix $K_{ab}$ is of the form
\begin{equation}
    K_{ab} = \begin{pmatrix}
        -i z & i k & 0 \\
        i \beta_1 k & \Gamma + \gamma_s k^2 - i z & i \beta_2 k \\
        \alpha_1 k^2 \sigma & ik \frac{n_0}{\varepsilon_0 + P_0} & \alpha_2 k^2 \sigma - i z
    \end{pmatrix},
\end{equation}
where we defined
\begin{align}
    \gamma_s &= \frac{1}{\varepsilon_0 + P_0} \left(\frac{2d-2}{d} \eta + \zeta\right),\\
    \alpha_1 &= \left(\frac{\partial \mu}{\partial \varepsilon}\right)_n -\frac{\mu_0}{T_0} \left(\frac{\partial T}{\partial \varepsilon}\right)_n,\quad \alpha_2 = \left(\frac{\partial \mu}{\partial n}\right)_\varepsilon -\frac{\mu_0}{T_0} \left(\frac{\partial T}{\partial n}\right)_\varepsilon,\\
    \beta_1 &= \left(\frac{\partial P}{\partial \varepsilon}\right)_n,\quad \beta_2 = \left(\frac{\partial P}{\partial n}\right)_\varepsilon.
\end{align}
The solution to the initial value problem \eqref{can-sol} is connected to the (canonical) retarded correlator via 
\begin{equation}\label{can-sol-corr}
    \varphi_a(z,k)= -\frac{1}{i z}\left( G^{\rm (can)}_{ab} (z,\,\textbf{k}) - G^{\rm (can)}_{ab} (z=0,\,\textbf{k}) \right)\lambda_b.
\end{equation}
Comparing \eqref{can-sol} and \eqref{can-sol-corr} we compute the expression for the canonical retarded correlator:
\begin{align}\label{eq:canonicalCorrelator}
    G^{\rm (can)}_{ab}(z,\textbf{k})&=-\left(\delta_{ac}+i z (K^{-1})_{ac}\right)\chi_{cb}.
\end{align} 
We note that the $G_{ab}$ matrix is symmetric. 
The decoupled perpendicular $\delta \pi^\perp$ correlators are obtained following the same procedure. In this way we gain access to the retarded correlators for conserved quantities\footnote{Even though $\delta \pi^i$ are not strictly speaking conserved we nevertheless have access to the $\delta \pi^i$ correlators due to the fact that $\delta \pi^i$ possess a well-defined initial value problem given in \eqref{eq:M-systemEq}.} $\delta \varepsilon$, $\delta \pi^i$ and $\delta n$.

Using the general expressions of the correlators obtained above, namely, inserting \eqref{eq:Kmatrix} and \eqref{eq:hydroChi} into \eqref{eq:canonicalCorrelator}, we obtain the following expressions for the correlators in general $d$: 
\begin{align}\label{eq:HydroCorrelators}
	G_{\varepsilon\varepsilon} &=\frac{(\varepsilon_0+P_0)k^2}{\mathcal{D}(\omega,k)}\left(\omega + i \alpha_2 \sigma k^2 \right),\\
	G_{\varepsilon\pi^\parallel} &= \frac{\omega}{k}G_{\varepsilon\varepsilon},\\
	G_{\varepsilon n} &= \frac{k^2}{\mathcal{D}(\omega,k)}\left(n_0\omega-i\alpha_1\sigma (\varepsilon_0+P_0)\,k^4\right),\\   
	G_{\pi^\parallel \pi^\parallel} &=\frac{k^2 \omega  \left(-i \gamma_s \omega +\alpha_2 \Gamma_\parallel \sigma +v_s^2 \right)-i\Gamma_\parallel \omega ^2-i k^4 \sigma 
		(\alpha_1 \beta_2-\alpha_2 \beta_1)}{\mathcal{D}(\omega,k)},\\
	G_{\pi^\parallel n} &= \frac{\omega}{k} G_{\varepsilon n},\\
    G_{n n} &=  \frac{1}{(\varepsilon_0+P_0)\beta_2\mathcal{D}(\omega,k)}\notag\\
    &\times \Big[k^2 \omega  \left(\alpha_2 n_0 \sigma  (\varepsilon_0+P_0) (\Gamma_\parallel-i \omega )+\alpha_1 \sigma  (\varepsilon_0+P_0)^2 (\Gamma_\parallel-i
	\omega )+\beta_2 n_0^2\right)\label{eq:HydroCorrelatorsF}\\
	&+k^4 \sigma  (\varepsilon_0+P_0) (n_0 (-i \alpha_1 \beta_2 +i \alpha_2
	\beta_1+\alpha_2 \gamma_s \omega )+\alpha_1 (\varepsilon_0+P_0) (\gamma_s \omega +i \beta_1))\Big]\notag,
\end{align}
where
\begin{align}
    \mathcal{D}(\omega,k)&= \omega^3 + i \omega^2 \left[\Gamma_\parallel + \left(\alpha_2 \sigma + \gamma_s\right)k^2\right] \notag\\
	&\quad- i k^4 \sigma \left(\alpha_2 \beta_1 - \alpha_1 \beta_2\right) - k^2 \omega\left(v_s^2 + \alpha_2 \sigma \Gamma_\parallel\right) + \mathcal{O}(k^4\omega)
\end{align}
and $v_s^2 \equiv \beta_1 + \beta_2 n_0/(\varepsilon_0+P_0)$. For the transverse channel, we obtain the damped diffusive correlator \cite{Davison:2014lua}
\begin{equation}
    G_{\pi^\perp \pi^\perp} = \frac{-(\varepsilon_0+P_0)\left(\frac{\eta}{\varepsilon_0 + P_0} k^2 + \Gamma_\perp\right)}{-i\omega + \frac{\eta}{\varepsilon_0 + P_0} k^2 + \Gamma_\perp}.
\end{equation}

Using the correlators \eqref{eq:HydroCorrelators}--\eqref{eq:HydroCorrelatorsF}, one can then calculate the \textit{longitudinal} thermoelectric conductivities as described in the main text. Indeed, using the Ward identities, this approach additionally yields only $G_{J_\parallel J_\parallel}$ and $G_{\pi_\parallel J_\parallel}$. This is one of the main advantages of the variational approach over the canonical setup. For the longitudinal conductivities we obtain:
\begin{align}\label{eq:HydroConductivity1}
    \sigma_{xx}^{\rm(hydro)} &= \sigma_Q^{\rm(hydro)} + \frac{n_0^2}{\varepsilon_0 + P_0} \frac{i}{\omega + i \Gamma},\\
    \alpha_{xx}^{\rm(hydro)} &= -\frac{\mu_0}{T_0}\sigma_Q^{\rm(hydro)} - \frac{s_0 n_0}{\varepsilon_0 + P_0} \frac{i}{\omega + i \Gamma},\\
    \bar{\kappa}_{xx}^{\rm(hydro)} &= \frac{\mu_0^2}{T_0}\sigma_Q^{\rm(hydro)} + \frac{s_0^2 T_0}{\varepsilon_0 + P_0} \frac{i}{\omega + i \Gamma}\label{eq:HydroConductivity3},
\end{align}
where
\begin{align}\label{eq:HydroSigmaQ}
    \sigma_Q^{\rm(hydro)} = \frac{\alpha_2 n_0 + \alpha_1 \left(\varepsilon_0+P_0\right)}{\beta_2}\sigma.
\end{align}
Notably, the Onsager relations hold as well as equations \eqref{eq:WardConductivities}.

The same procedure can be executed using the $\delta J^\mu$ and $\delta T^\mn$ obtained by taking first and second moments of the solution to the linearized Boltzmann equation \eqref{bolt-sol}, respectively. We note that the canonical kinetic theory correlators obtained this way agree with the ones in the main text up to contact terms.

{We highlight that in the massless kinetic case presented in the main text, $\beta_1 = 1/d$ and $\beta_2=0$.\footnote{Since $\alpha_1$ and $\alpha_2$ do not play a significant role in this discussion we leave them in a general form. One can show that our equation of state corresponds to $\alpha_1 = -1/n_0$ and $\alpha_2 = (d+1)/\chi$.} Crucially, this means that the denominator structure factorizes, leading to a cancellation of the diffusion pole in the energy-energy correlator at finite density\footnote{We thank Blaise Gout{\' e}raux for discussions on this point.}:
\begin{align}
G_{\varepsilon\varepsilon} &=\frac{-d(\varepsilon_0+P_0) ik^2}{d(\Gamma_\parallel - i \omega)\omega + (d\gamma_s \omega + i)k^2}.
\end{align}
Furthermore, the $\langle TT \rangle$ correlators are unaffected by the non-zero $\mu_0$, as they agree identically with the $\mu_0 = 0$ calculation (cf.~Ref.~\cite{Davison:2014lua}). This again agrees with the kinetic theory results presented in the main text. Repeating the above reasoning we again conclude that this is a consequence of the form of our equation of state.

We note that the denominator of the $G_{nn}$ correlator factorizes into a structure with diffusion and sound modes:\footnote{The $\mu_0 = 0$ results can be found in \cite{Kovtun:2012rj}.}
\begin{align}
    G_{nn} &= \frac{d\left(\alpha_1 n_0 - \alpha_2 \sigma (\Gamma_\parallel - i \omega)\right)k^2\omega - i \alpha_1^2 d(\varepsilon_0+P_0) \sigma k^4-\alpha_2 \sigma(2\gamma_s \omega +i)k^4 }{\alpha_2\left(\alpha_2\sigma k^2 - i \omega\right)\left(d(\Gamma_\parallel - i \omega)\omega + (2\gamma_s\omega + i)k^2\right)}.
\end{align}This is analogous to our results in kinetic theory, as noted in the main text. Since these two properties are not present generally (see~\eqref{eq:HydroCorrelators}--\eqref{eq:HydroCorrelatorsF}), we conclude that they are a consequence of our particular equation of state that arises from the kinetic theory setup.

Also, we note that only difference in conductivities \eqref{eq:HydroConductivity1}--\eqref{eq:HydroConductivity3} is in the term $\sigma_Q$ \eqref{eq:HydroSigmaQ} which is now simply $\sigma_Q = \sigma$.}

\section{Contact terms}\label{app:contact}
{Here, we provide further technical details on the relevant contact terms that enter the kinetic theory calculation.} We begin by analyzing the current conservation law which is in the first order of the form
\begin{align}
    \partial_\mu \delta J^\mu &= \int \frac{d^d \textbf{p}}{(2\pi)^d p^0}p^\mu \partial_\mu \delta f \notag\\
	&= \int \frac{d^d \textbf{p}}{(2\pi)^d p^0} \frac{f_0}{T_0}  \left(- \Gamma^0_\ab  p^\alpha p^\beta + E^i p_i\right)=-\Gamma^0_\ab \frac{T^\ab_0}{T_0},
\end{align}
where we used the Boltzmann equation \eqref{eq:Boltzmann} and assumed the matching conditions \eqref{matching} hold. {Taking the functional derivative with respect to $\delta A_\mu$ and $\delta g_\mn$, we obtain the following relations in the Fourier space:}
\begin{equation}\label{eq:Ward-J}
	ik_\mu G^{\mu,\nu}_{JJ} = 0, \quad ik_\mu G^{\mu,\ab}_{JT} = 2 \frac{T^\mn_\mathrm{eq}}{T_0} \frac{\delta \Gamma^0_\mn}{\delta g_\ab}.
\end{equation}
{We can proceed analogously for the energy-momentum conservation:}
\begin{align}
	\partial_\mu \delta T^\mn &= \int \frac{d^d \textbf{p}}{(2\pi)^d p^0}p^\nu p^\mu \partial_\mu \delta f \notag \\
	&= \int \frac{d^d \textbf{p}}{(2\pi)^d p^0} \frac{f^{(0)}_{\mathrm{eq}}}{T_0} p^\nu  \left(- \Gamma^0_\ab  p^\alpha p^\beta + E^i p_i - p^0 p^j \Gamma_{ij}\left(\delta u^i - \delta g_{0k}\delta^{ki}\right)\right) \notag \\
	&=  d\,(\varepsilon_0 + P_0)\left[-\Gamma^0_\ab I^{\alpha \beta \nu} - I^{\nu 0 j}\Gamma_{ij}\left(\delta u^i - \delta g_{0k}\delta^{ki}\right)\right] + \frac{1}{T_0} E_i T^{i\nu}_0,
\end{align}
where we defined
\begin{equation}\label{eq:Iabc}
	I^{\alpha\beta\nu} \equiv \int \frac{d \Omega^{(d)}}{\Omega^{(d)}} v^\alpha v^\beta v^\nu
\end{equation}
and used the definitions of \eqref{eq:enth}. Note that the index symmetry implies $I^{\alpha\beta\nu} = \delta^{(\alpha}_0 \delta^{\beta\nu)}$ and, in particular, that $I^{0ij} = \delta^{ij}/d$. Consequently, we obtain the Ward identities for the $\langle TT \rangle$ and $\langle TJ \rangle$ correlators:
\begin{align}\label{eq:Ward-T1}
	&i k_\mu G^{\mu\hat{\nu},\alpha}_{TJ}+ \Gamma_{\hat{\nu}}G^{0\hat{\nu},\alpha}_{TJ} = -\frac{1}{T_0} T^{j\hat{\nu}}_0 \frac{\delta E_j}{\delta A_\alpha},\\
	&ik_\mu G^{\mu \hat{\nu}, \alpha\beta}_{TT} + \Gamma_{\hat{\nu}}G^{0\hat{\nu},\alpha\beta}_{TT} = 2 d\,(\varepsilon_0 + P_0)\left( I^{\rho\sigma \hat{\nu}} \frac{\delta \Gamma^0_{\rho\sigma}}{\delta g_\ab} -   \frac{1}{d} \Gamma_{\hat{\nu}}\delta^{\hat{\nu}k} \frac{\delta g_{0k}}{\delta g_\ab} \right),\label{eq:Ward-T2}
\end{align}
where the parameters $\Gamma_\nu$ are the diagonal components of the object $\Gamma_\mn$ defined in the main text and the hatted indices are not summed over. We note that the first identity \eqref{eq:Ward-T1} could equivalently be obtained by considering $\partial_\mu \delta T^\mn = F^{\nu\alpha}J_\alpha$.

\bibliography{literature}
\bibliographystyle{JHEP}

\end{document}